\documentclass[11pt,a4paper]{article}

\usepackage{amsmath, amssymb, amsthm}
\usepackage[title]{appendix}
\usepackage{graphicx}
\usepackage[numbers,square]{natbib}
\usepackage{geometry}
\usepackage{hyperref}
\usepackage{xcolor}
\usepackage{float}

\title{Libby-Fox perturbations and the semi-analytic adjoint solution for laminar viscous flow along a flat plate}
\author{Carlos Lozano and Jorge Ponsin  \\
Theoretical and Computational Aerodynamics  \\
National Institute of Aerospace Technology (INTA), Spain  \\
*Corresponding author: lozanorc@inta.es }
\date{\today}

\begin{document}

\maketitle

\begin{abstract}

The properties of the solution to the adjoint two-dimensional boundary layer (BL) equations on a flat plate are investigated from the viewpoint of Libby-Fox theory, which describes the algebraic perturbations to the Blasius boundary layer. 
The adjoint solution is obtained from the Green's function of the perturbation equation as a sum over the infinite perturbation modes of the Blasius solution. The explicit representation of the adjoint solution allows us to derive constraints on the eigenvalues and eigenfunctions, explicitly compute the Adjoint Transport Convection (ATC) term and evaluate flow sensitivities for shape design, initial-value perturbations, and active flow control. The extension of the analysis to the case with non-zero pressure gradient, corresponding to the Falkner-Skan solution, is also briefly discussed. 

\end{abstract}

\section{Introduction}
\label{sec:introduction}

The Blasius solution is one of the most remarkable solutions in fluid mechanics. 
It describes the steady, two-dimensional laminar boundary layer that forms over a semi-infinite flat plate held parallel to a constant unidirectional flow. 
While surprisingly accurate for many engineering purposes, the Blasius solution is only an approximate (first-order) solution, as it predicts a non-zero normal velocity at the outer edge of the boundary layer, which is physically inconsistent with the outer flow assumptions. 
In spite of this shortcoming, the Blasius solution remains a valuable asset as a benchmark for validating numerical simulations of boundary layer flows and as the base flow profile for boundary layer stability analysis. 
It must be emphasized that the steady Blasius boundary layer is physically unstable via different mechanisms such as Tollmien-Schlichting waves; the present work does not attempt to predict transition, but rather focuses strictly on the exact mathematical structure of the steady adjoint boundary layer equations. 

Linearized perturbations to a Blasius boundary layer have been considered in many contexts. The most relevant to us is the work of Libby \& Fox \cite{LibbyFox1963}, who identified an infinite tower of two-dimensional (steady) perturbation modes and obtained the discrete eigenvalues and eigenfunctions of the perturbation equation and applied them to the analysis of first-order boundary layer solutions deviating slightly from the Blasius solution. Subsequent work clarified the mathematical problem, provided more eigenvalues and derived approximate and asymptotic formulae for the eigenvalues. A comprehensive review of those early developments can be found in \cite{Libby1970}. More recent work has generalized the Libby-Fox eigenmodes to more complex scenarios \cite{Luchini1996, Luchini2000, HewittDuck2018}. 

One of the interesting features of Libby and Fox's analysis is that it is possible to develop a Green's function solution that will enable us to directly link the perturbative and adjoint approaches.
%
%
Adjoint equations have many applications across mathematics, engineering, and the sciences. 
Of particular interest to us are applications related to adjoint-based sensitivity analysis \cite{PeterDwight2010}, where adjoint techniques are highly efficient in determining the sensitivity of output quantities (such as lift and drag) to a large number of input data variables. 
This interpretation is at the heart of most engineering applications of adjoint techniques in fluid dynamics, including flow stability \cite{LuchiniBottaro2014}, aerodynamic design \cite{Jameson1988, GilesPierce2000}, flow control \cite{Lions1971, LiuMacArt2024}, and error estimation \cite{FidkowskiDarmofal2011}. 
Another interpretation, more technical in nature but directly related to the previous one, will be of primary concern here: the adjoint variable at a specific location is precisely the value of the linearized objective function evaluated with the appropriate Green's function of the original problem \cite{GilesPierce2001}. This link with the Green's function has been used to obtain exact adjoint solutions \cite{ LozanoPonsin2022} and will be exploited here for the analysis of the adjoint solution to the boundary layer equations. 

Adjoint methods have been applied to boundary layer flows in a variety of contexts. The adjoint of the Navier-Stokes, parabolized stability, Orr-Sommerfeld or boundary layer equations have been used in shape optimization, flow control, boundary layer stability, adjoint-based global-mode stabilization, boundary layer receptivity, structural sensitivity, identification of optimal perturbations and resolvent analysis. The literature is too vast to be properly summarized here, but \cite{LuchiniBottaro2014} contains a comprehensive overview of adjoint applications to hydrodynamic stability up to 2014.

Directly related to the present paper is the recent work \cite{Kuhl2021}, where a thorough analysis of the complete adjoint boundary–layer equations and dynamics is presented, including a  proposal for an analytic adjoint solution for the Blasius boundary 
layer. It also contains the intriguing idea that the adjoint solution develops a boundary layer structure 
that emanates from the end of the plate, as corresponds to the reversed convective characteristics 
relative to the primal flow. Their analysis differs from ours in several aspects and their proposed 
solution appears to be in conflict with the solution that we will present below except under special 
circumstances.

The present paper explicitly casts the solution to the adjoint of the Prandtl boundary layer equations on a flat plate as a Libby-Fox eigenfunction expansion and discusses its convergence properties. The restricted scope of the paper poses a clear limitation of this work, its direct relevance to more complex geometries, unsteady perturbations or fully turbulent flows being debatable. On the other hand, an improved 
understanding of the adjoint solution corresponding to Blasius boundary layer is relevant in several 
regards. First, even though assembling the solution requires significant numerical work, it can be used as a validation case for adjoint Navier-Stokes or boundary layer solvers and can also help guide the proper formulation of the adjoint problem and its boundary conditions (b.c.) in numerical implementations. The solution can also be used to understand the mathematical structure of the 
adjoint problem in a simplified, yet fundamental, case.  

Likewise, an analytic solution facilitates theoretical developments, such as deriving constraints on the eigenvalues and eigenfunctions of the primal problem, and analyzing the Adjoint Transport Convection (ATC) term, whose actual value is hard to determine from the viewpoint of original partial differential equations. Finally, an explicit adjoint solution can be used to directly compute gradients and sensitivities with respect to various control parameters such as initial velocity profiles, plate shape, and wall normal velocity distributions. 

To make the subject as accessible as possible, the paper is organized as follows.
Section \ref{sec:perturbations} reviews some basic facts about the Libby-Fox analysis of perturbations to the Blasius solution. 
Section \ref{sec:adjoint_formulation_orig} introduces the adjoint equations and derives the adjoint eigenvalue problem.
Section \ref{sec:analytic_solution} applies the Green's function approach to compute the semi-analytic adjoint solution. The solution is shown to obey the adjoint boundary conditions and the possibility that it can be written in a simple similarity form is explicitly refuted. 
Section \ref{sec:numerical_analysis} presents some numerical experiments including sample solutions evaluated from numerical integration of the adjoint Libby-Fox equation and 
presents empirical convergence properties and comparisons with numerical solutions obtained by discretizing the adjoint equations directly.
Section \ref{sec:applications} applies the analytic solution to the analysis of the ATC and to the sensitivity analysis of the drag coefficient in initial-value, shape design and active flow control (via blowing-suction) problems.
Section \ref{sec:falkner_skan} briefly examines the extension to Falkner-Skan flows. 
Section \ref{sec:conclusions} summarizes the findings of the paper and outlines future research. 

%
%

\section{Perturbations to the boundary layer equations}
\label{sec:perturbations}

This section contains a brief review of some well-known facts about linearized perturbations to the Blasius boundary layer solution as originally developed by \cite{LibbyFox1963}. There is much more to say about this topic than can be found here. Aside from the original reference, \cite{Libby1970} contains an excellent review, and we also found the introduction in \cite{Luchini1996} particularly useful.

The equations for the incompressible boundary layer along a flat plate with zero pressure gradient are:
\begin{align}
    u u_x + v u_y - \nu u_{yy} &= 0 \nonumber \\
    u_x + v_y &= 0 \label{eq:NS}
\end{align}
where $(x,y)$ are Cartesian coordinates chosen such that the plate extends along the positive $x$ axis, $(u,v)$ are the Cartesian components of the velocity and $\nu$ is the constant (kinematic) viscosity. In  (\ref{eq:NS}) and in the remainder of the paper, we will denote derivatives by subscripts so that $u_y = \partial_y u$. Equation (\ref{eq:NS}) can be equivalently written in terms of a streamfunction $\psi(x,y)$ as:
\begin{equation}
    \psi_y \psi_{xy} - \psi_x \psi_{yy} - \nu \psi_{yyy}=0 \label{eq:streamfunction}
\end{equation}
where $\psi(x,y)$ is related to the velocity as $(u,v) = (\psi_y, -\psi_x)$. 

An equivalent equation describing the boundary layer can be obtained by changing variables to $\psi(x,y) = \sqrt{2 Re_m \nu^2 x} F(x,\eta)$, where $\eta = y \sqrt{Re_m / (2x)}$ is Blasius similarity variable, $Re_m = \rho U / \mu = U / \nu$ is the Reynolds number per unit length (Reynolds per meter), and $\mu$, $\rho$ and $U$ are the viscosity, density and velocity at the edge of the boundary layer, respectively. With this definitions, (\ref{eq:streamfunction}) is transformed into  
\begin{equation}
    F_{\eta\eta\eta}(x,\eta) + F(x,\eta) F_{\eta\eta}(x,\eta) = 2x \left(F_\eta(x,\eta) F_{x\eta}(x,\eta) - F_{\eta\eta}(x,\eta) F_x(x,\eta)\right) \label{eq:nonsimilar}
\end{equation}
where now $x$ and $\eta$ have to be understood as formally independent variables. Equation (\ref{eq:nonsimilar}) admits the particular solution $ F(x,\eta) = F_0(\eta)$, where  $F_0$ is Blasius' similarity function obeying: 
\begin{equation}
    F_{0,\eta\eta\eta}(\eta) + F_0(\eta) F_{0,\eta\eta}(\eta) = 0 \label{eq:blasius}
\end{equation}
with $F_0(0) = 0$, $F_{0,\eta}(0) = 0$ and $F_{0,\eta}(\infty) = 1$. Libby and Fox investigated non-similar solutions to (\ref{eq:nonsimilar}) of the form $F(x,\eta) = F_0(\eta) + \delta F(x,\eta)$ where $\delta F$ is a small perturbation. The equation for $\delta F$ is obtained by linearizing  (\ref{eq:nonsimilar}) about $F_0$ and is
\begin{equation}
    \delta F_{\eta\eta\eta} + F_0 \delta F_{\eta\eta} + F_{0,\eta\eta} \delta F + 2x (F_{0,\eta\eta} \delta F_x - F_{0,\eta} \delta F_{x\eta}) = 0 \label{eq:linearized_perturbation}
\end{equation}
Equation (\ref{eq:linearized_perturbation}) has separable solutions of the form:
\begin{equation}
    \delta F(x,\eta) = x^{-\lambda/2} N(\eta) \label{eq:separation}
\end{equation}
where the functions $N$ are solutions to the eigenvalue problem:
\begin{equation}
    N_{\eta\eta\eta} + F_0 N_{\eta\eta}+ \lambda F_{0,\eta} N_\eta  +(1- \lambda) F_{0,\eta\eta} N = 0 \label{eq:eigenvalue_problem}
\end{equation}
with the conditions $N(0)=0$, $N_\eta(0)=0$, $N_\eta(\infty)=0$ as well as the scaling condition $N_{\eta\eta}(0)=1$. 

Imposing the additional condition that $N_\eta$ vanishes exponentially fast at infinity allowed Libby and Fox to transform the eigenvalue problem (\ref{eq:eigenvalue_problem}) into a second-order Sturm-Liouville equation:
\begin{equation}
    \left( \frac{F_{0,\eta}^3}{F_{0,\eta\eta}} H_{\eta} \right)_\eta + \lambda \frac{F_{0,\eta}^4}{F_{0,\eta\eta}} H - F_0 F_{0,\eta}^2 H = 0 \label{eq:sturm_liouville}
\end{equation}
where $H = (N/F_{0,\eta})_\eta$. Consequently, the eigenvalues form an infinite discrete sequence of positive real numbers $\lambda_k > 0$, $k=1,2,...$ and the eigenfunctions form a complete orthogonal set obeying:
\begin{equation}
    \int_0^\infty \frac{F_{0,\eta}^4}{F_{0,\eta\eta}} \left(\frac{N_k}{F_{0,\eta}}\right)_\eta \left(\frac{N_j}{F_{0,\eta}}\right)_\eta d\eta = \delta_{kj} C_k \label{eq:orthogonality}
\end{equation}
and can be used to represent any physically admissible perturbation to the Blasius boundary layer. 

The first eigenvalue can be determined exactly as $\lambda_1 = 2$ and its corresponding eigenfunction is:
\begin{equation}
    N_1 = (\eta F_{0,\eta} - F_0)/\alpha \label{eq:first_eigenfunction}
\end{equation}
where $\alpha = F_{0,\eta\eta}(0)$, such that $N_1/x$ is a solution to  (\ref{eq:linearized_perturbation}) that corresponds to a simple translation of the origin of coordinates \cite{Stewartson1957}. The higher ($k>1$) eigenvalues and eigenfunctions are hard to obtain and are only amenable to numerical treatment. The next few eigenvalues are 3.774, 5.629, 7.513, so that the general solution has the form:
\begin{equation}
\begin{aligned}
    \delta F(x,\eta) &= \sum_{k=1}^\infty A_k x^{-\lambda_k/2} N_k(\eta) \\
    &= A_1 x^{-1} (\eta F_{0,\eta} - F_0)/\alpha + A_2 x^{-1.887} N_2 + A_3 x^{-2.814} N_3 + A_4 x^{-3.757} N_4 + \dots
\end{aligned} \label{eq:general_solution}
\end{equation}

The eigenvalue problem  (\ref{eq:eigenvalue_problem}) is singular in that the boundary conditions, alone, are not sufficient to determine the spectrum. Determination of the spectrum requires imposing on the eigenfunctions an exponential decay at infinity. A variety of techniques has been used to determine the eigenvalues and eigenfunctions of  (\ref{eq:eigenvalue_problem}) (see \cite{Libby1970} for a review of early developments). The original work \cite{LibbyFox1963} only gave the first 10 eigenvalues. Libby \cite{Libby1965}, Wilks and Bramley \cite{WilksBramley1977} and Luchini \cite{Luchini1996} listed the first twenty eigenvalues. These are all exact approaches. Libby and Fox's and Luchini's approaches are based on the matching of a forward integration ($\eta$ increasing) of the full equation (\ref{eq:eigenvalue_problem}) with a backward integration ($\eta$ decreasing) of the asymptotic form of  (\ref{eq:eigenvalue_problem}), while Wilks and Bramley's is based on a Ricatti transformation of the perturbation equations \cite{Wilks1979}. 

On the other hand, Kotorynski \cite{Kotorynski1968} derived approximate values for the eigenvalues replacing the coefficient functions of  (\ref{eq:eigenvalue_problem}) by the sum of its asymptotic terms near the wall and far from the wall. Kemp \cite{Kemp1970} improved Kotorynski's formula, giving approximate values for the norms as well and, additionally, applied the results of a WKB approximation to obtain a more accurate approximation for the eigenvalues. Finally, Brown \cite{Brown1968}, using matched asymptotic expansions, obtained the following asymptotic formula for the eigenvalues:
\begin{equation}
    \lambda_k = 2(k-1) - 0.2705\sqrt{2(k-1)} + 2.203 - \frac{0.076}{\sqrt{2(k-1)}} + \mathcal{O}\left(\frac{1}{2(k-1)}\right) \label{eq:brown_formula}
\end{equation}
(valid for $k>1$) which works surprisingly well, yielding a relative error of 0.18\% already at $k=2$.

\section{Adjoint boundary-layer problem formulation}
\label{sec:adjoint_formulation_orig}

There are several ways to define the adjoint equations associated with the Blasius solution.
One can consider the adjoint to the full Navier-Stokes equations about the Blasius solution, the adjoint to the boundary layer equations, or, finally, the adjoint to Prandtl's boundary layer equations for the flat plate \eqref{eq:NS}. We will consider the 
last option. The corresponding derivation of the adjoint equations and boundary conditions has been 
discussed at length in \cite{Kuhl2021} and will not be repeated here. 

%
%

The application of the adjoint approach in sensitivity analysis is based on the premise that one wishes to compute the sensitivity of a certain quantity of interest or objective function to flow perturbations. This can be done directly by solving the linearized flow equations. However, if the number of perturbations is large (as in, for example, a shape design problem with many design variables) the cost soon becomes prohibitive. The adjoint approach is designed to bypass the need to compute the linearized flow equations for each independent perturbation, requiring instead the solution of a single partial differential equation for each independent objective function.    

In the present case, the selected objective function is the integrated friction drag coefficient along a section of length $L$ of the plate, which is defined as:
\begin{equation}
    C_D(L) = \tilde{\kappa} \int_0^L (u_y)_{y=0} dx \label{eq:drag_coefficient}
\end{equation} 
where $\tilde{\kappa} = 2\mu/(\rho U^2 L) = 2/(Re_m L U)$ is a normalization constant.
Rewriting  \eqref{eq:drag_coefficient} in terms of the stream function and using a Lagrange multiplier $\tilde{\psi}$ to enforce   the flow equation \eqref{eq:streamfunction} yields the Lagrangian:
\begin{equation}
    \mathcal{L} = \tilde{\kappa} \int_0^L (\psi_{yy})_{y=0} dx - \int_0^L \int_0^\infty \tilde{\psi} (\psi_y \psi_{xy} - \psi_x \psi_{yy} - \nu \psi_{yyy}) dx dy \label{eq:lagrangian}
\end{equation} 
Equation \eqref{eq:lagrangian} sets the basis for the adjoint treatment and also helps to fix notation as well as the normalization of the adjoint field $\tilde{\psi}$.  
Notice that, by using the streamfunction formulation, there is no need to add to  \eqref{eq:lagrangian} a term to enforce the incompressibility condition. 

Linearizing  \eqref{eq:lagrangian} with respect to flow perturbations $\delta\psi$, integrating by parts and rearranging yields:
\begin{equation}
\begin{aligned}
    \delta\mathcal{L} = \delta C_D(L) &= \tilde{\kappa} \int_0^L (\delta\psi_{yy})_{y=0} dx \\
    &\quad - \int_0^L \int_0^\infty \delta\psi (\psi_y \tilde{\psi}_{xy} - \psi_x \tilde{\psi}_{yy} - 2\psi_{xy} \tilde{\psi}_y + 2\psi_{yy} \tilde{\psi}_x + \nu \tilde{\psi}_{yyy}) dx dy \\
    &\quad + \int_0^L dx \Big[ \tilde{\psi} \psi_x \delta\psi_y - 2\tilde{\psi} \psi_{xy} \delta\psi - \tilde{\psi}_y \psi_x \delta\psi - \tilde{\psi} \psi_y \delta\psi_x\Big]_{y=0}^{y=\infty} \\
    &\quad + \int_0^L dx \Big[  \nu \tilde{\psi} \delta\psi_{yy} - \nu \tilde{\psi}_y \delta\psi_y + \nu \tilde{\psi}_{yy} \delta\psi \Big]_{y=0}^{y=\infty} \\
    &\quad + \int_0^\infty dy \Big[ \tilde{\psi}_y \psi_y \delta\psi + 2\tilde{\psi} \psi_{yy} \delta\psi \Big]_{x=0}^{x=L}
\end{aligned} \label{eq:linearized_lagrangian}
\end{equation} 
For design applications,  \eqref{eq:linearized_lagrangian} contains an additional term that corresponds to the perturbation of the plate geometry.
This term does not play any role in the derivation of the adjoint equations and is not included here, but it will be briefly discussed in section \ref{sec:shape_sens}.

The adjoint problem is defined to eliminate the dependence of  \eqref{eq:linearized_lagrangian} on $\delta\psi$ and its derivatives so that the sensitivity of the cost function to flow perturbations can be computed without having to solve the linearized perturbation equation. Vanishing of the domain integral requires that $\tilde{\psi}$ obeys the adjoint equation:
\begin{equation}
    \psi_y \tilde{\psi}_{xy} - \psi_x \tilde{\psi}_{yy} - 2\psi_{xy} \tilde{\psi}_y + 2\psi_{yy} \tilde{\psi}_x + \nu \tilde{\psi}_{yyy} = 0 \label{eq:adjoint_streamfunction}
\end{equation}

The boundary terms are tackled in a similar fashion. In some of them, $\delta\psi$ and its derivatives cannot be computed without solving the linearized flow equations. Others are completely specified by the linearized boundary conditions. The former are eliminated by imposing the following boundary conditions on the adjoint variable:
\begin{equation}
\begin{aligned}
    \tilde{\psi}(x,0) &= \frac{\tilde{\kappa}}{\nu} = \frac{2}{L U^2} \\
    \tilde{\psi}(L,y) &= 0 \\
    \tilde{\psi}(x,\infty) &= 0
\end{aligned} \label{eq:adjoint_boundary_conditions}
\end{equation}
The remaining boundary terms yield the sensitivity derivatives. We will use some of them in section \ref{sec:applications}.

It may be useful at this stage to compare  \eqref{eq:adjoint_streamfunction} to more conventional adjoint treatments.
Equation \eqref{eq:adjoint_streamfunction} is a single, third-order equation for one adjoint variable, when in view of  \eqref{eq:NS} one would expect two adjoint variables obeying two equations of second order at most. 
In fact, the adjoint equations corresponding to  \eqref{eq:NS} can be easily shown to be 
\begin{align}
    \tilde{\psi}_{p,x} + u \tilde{\psi}_x + v \tilde{\psi}_y - \tilde{\psi} u_x + \nu \tilde{\psi}_{yy} &= 0 \nonumber \\
    \tilde{\psi}_{p,y} - \tilde{\psi} u_y &= 0 \label{eq:adjoint_eq}
\end{align} 
where $\tilde{\psi}_p$ is the adjoint variable (the Lagrange multiplier) that corresponds to the continuity equation (usually known as the adjoint pressure). 
$\tilde{\psi}_p$ can be eliminated from \eqref{eq:adjoint_eq} by applying $\partial_y$ to the first equation, $\partial_x$ to the second, and subtracting the second equation from the first.
The resulting equation, after setting $(u,v) = (\psi_y, -\psi_x)$, is precisely  \eqref{eq:adjoint_streamfunction}.
Once  \eqref{eq:adjoint_streamfunction} is solved, $\tilde{\psi}_p$ can be obtained from  \eqref{eq:adjoint_eq}. 

We now change variables to $(x,\eta)$ and write $\psi(x,y) = \sqrt{2 Re_m \nu^2 x} F(x,\eta)$.
In the new variables the drag can be computed as:
\begin{equation}
    C_D(L) = \kappa \int_0^L \frac{1}{\sqrt{2x}} F_{\eta\eta}(x,0) dx \label{eq:drag_eta}
\end{equation} 
where $\kappa = 2/(\sqrt{Re_m} L)$, and the penalty term in  \eqref{eq:lagrangian} gets converted to:
\begin{equation}
    \int_0^L \int_0^\infty \tilde{\psi} (\psi_y \psi_{xy} - \psi_x \psi_{yy} - \nu \psi_{yyy}) dx dy = \iint dx d\eta \tilde{\Psi} \big( F_{\eta\eta\eta} + F F_{\eta\eta} - 2x(F_\eta F_{x\eta} - F_{\eta\eta} F_x) \big) \label{eq:penalty_eta}
\end{equation}
where
\begin{equation}
    \tilde{\Psi}(x,\eta) = - \frac{U^2}{\sqrt{2 Re_m x}} \tilde{\psi}(x,\eta) \label{eq:adjoint_variable_transformation}
\end{equation} 
is the adjoint variable in the new formulation. Setting $F = F_0 + \delta F$ in \eqref{eq:penalty_eta}, where $F_0(\eta)$ is Blasius solution, linearizing with respect to the small perturbation $\delta F$ and integrating by parts in $x$ and $\eta$ yields the following adjoint equation:
\begin{equation}
    -\tilde{\Psi}_{\eta\eta\eta} + F_0 \tilde{\Psi}_{\eta\eta} - 2x F_{0,\eta} \tilde{\Psi}_{x\eta} - 4x F_{0,\eta\eta} \tilde{\Psi}_x - 2F_{0,\eta\eta} \tilde{\Psi} = 0 \label{eq:linearized_adjoint_eta}
\end{equation} 
with boundary conditions:
\begin{equation}
\begin{aligned}
    \tilde{\Psi}(x,0) &= - \frac{\kappa}{\sqrt{2x}} \\
    \tilde{\Psi}(L,\eta) &= 0 \\
    \tilde{\Psi}(x,\infty) &= 0
\end{aligned} \label{eq:adjoint_eta_boundary_conditions}
\end{equation}

Consider the solution for $\tilde{\Psi}(x,\eta)$. Separation of variables so that $\tilde{\Psi}(x,\eta) = x^{\lambda/2-1}D(\eta)$ yields:
\begin{equation}
    -D_{\eta\eta\eta} + F_0 D_{\eta\eta} + (2 - \lambda) F_{0,\eta} D_\eta + 2(1 - \lambda) F_{0,\eta\eta} D = 0 \label{eq:adjoint_eigenvalue_problem}
\end{equation} 
which happens to be adjoint to the flow eigenvalue problem  \eqref{eq:eigenvalue_problem}. The solution to (\ref{eq:adjoint_eigenvalue_problem}) can be found by exploiting the underlying Sturm-Liouville structure of the Libby-Fox problem. As shown in Section \ref{sec:perturbations}, the third-order primal equation for $N$ \eqref{eq:eigenvalue_problem} can be transformed into a Sturm-Liouville equation (\ref{eq:sturm_liouville}) for the auxiliary variable $H(\eta)$. Both equations are related by an overall weight $W=F_{0,\eta\eta}/F_{0,\eta}^2$. Since the Sturm-Liouville equation is formally self-adjoint, a standard continuous adjoint argument shows that if $H_k$ is an eigensolution with eigenvalue $\lambda_k$, then $D_k(\eta)=H_k/W$ is an eigensolution of the adjoint equation with the same eigenvalue. Hence,
\begin{equation}
D_k(\eta) = \frac{1}{C_k} \frac{F_{0,\eta}^2}{F_{0,\eta\eta}} H_k(\eta)
     = \frac{1}{C_k} \left( \frac{F_{0,\eta}}{F_{0,\eta\eta}} N_{k,\eta} - N_k \right)
     \label{eq:D_k_H_k_relation}
\end{equation} 
where $C_k$ (the square of the norm of the $k^{th}$ Libby-Fox eigenfunction) has been inserted for future convenience. Using the explicit representation given by  \eqref{eq:D_k_H_k_relation}, it is easy to see that $D_k$ obeys the third-order equation \eqref{eq:adjoint_eigenvalue_problem} for $\lambda=\lambda_k$. We can also deduce that the adjoint eigenfunctions diverge polynomially as $\eta \rightarrow \infty$ (see appendix \ref{app:asymptotic_large_eta}), while at the wall they obey the boundary conditions $D_k(0) = 0$, $D_{k,\eta}(0) = 0$, $D_{k,\eta\eta}(0) = 1/C_k$. 

 Likewise, substituting (\ref{eq:D_k_H_k_relation}) into equation (\ref{eq:sturm_liouville}), it is straightforward to verify that the adjoint eigenfunctions $D_k$ obey the following Sturm-Liouville equation:
\begin{equation}
    \left( \frac{F_{0,\eta\eta}}{F_{0,\eta}} D_{k,\eta} \right)_\eta + (\lambda_k - 1) F_{0,\eta\eta} D_k = 0 \label{eq:adjoint_sturm_liouville}
\end{equation} 
Hence, the adjoint eigenfunctions form a complete set and are mutually orthogonal with respect to the weight $F_{0,\eta\eta}$:
\begin{equation}
    \int_0^\infty F_{0,\eta\eta} D_k D_j d\eta = C_k^{-1} \delta_{kj} \label{eq:adjoint_orthogonality}
\end{equation} 
where the normalization constant is inherited directly from  \eqref{eq:orthogonality}. Likewise, the full adjoint solution can be represented as an infinite sum over $D_k$. The coefficients of the modes must be fixed from the adjoint boundary conditions, which may prove to be a formidable task. Instead, we will determine them in the following section using the Green’s function approach.

\section{Analytic adjoint solution from the Green's function approach}
\label{sec:analytic_solution}

It is a well-established fact that the adjoint variable at a particular point is equal to the value of the corresponding linearized cost function evaluated using the Green's function for the same point \cite{GilesPierce2001}. Libby and Fox \cite{LibbyFox1963}  found that the Green's function that corresponds to  \eqref{eq:linearized_perturbation} is 
\begin{equation}
G(x',\eta'; x,\eta) = -{\cal H}(x'-x) \sum_{k=1}^\infty \frac{D_k(\eta)}{2x} \left(\frac{x'}{x}\right)^{-\lambda_k/2} N_k(\eta') \label{eq:libby_fox_green}
\end{equation} 
where ${\cal H}$ is the Heaviside function. As explained above, the adjoint variable $\tilde{\Psi}$ at $(x,\eta)$ is equal to the linearized drag associated to the Green's function \eqref{eq:libby_fox_green}. Using  \eqref{eq:drag_eta} we have
\begin{equation}
    \tilde{\Psi}(x,\eta) =\int_0^L \frac{\kappa}{\sqrt{2x'}} G_{\eta'\eta'}(x',\eta'; x,\eta)_{\eta'=0} dx' \label{eq:linearized_drag_green}
\end{equation} 
Inserting  \eqref{eq:libby_fox_green} into  \eqref{eq:linearized_drag_green}, integrating over $x$ and rearranging yields
%
%
\begin{equation}
    \tilde{\Psi}(x,\eta) = \frac{\kappa}{\sqrt{2x}} (\varphi(\xi,\eta) - 1) \label{eq:analytic_adjoint_drag}
\end{equation} 
where
\begin{equation}
    \varphi(\xi,\eta) = \sum_{k=1}^\infty \frac{D_k(\eta)}{\lambda_k - 1} \xi^{\frac{\lambda_k - 1}{2}} \label{eq:phi_definition}
\end{equation} 
and $\xi = x/L$. In the derivation of  \eqref{eq:analytic_adjoint_drag} we have used that
\begin{equation}
    \sum_{k=1}^\infty \frac{D_k(\eta)}{\lambda_k - 1} =1 \label{eq:D_0_definition}
\end{equation} 
A proof of  (\ref{eq:D_0_definition}) can be found in appendix \ref{app:D_0_is_1}. Equation \eqref{eq:analytic_adjoint_drag} is the drag-based adjoint solution for the equation \eqref{eq:linearized_perturbation} describing linearized perturbations to a Blasius base state.
The corresponding $x$-momentum adjoint solution for the boundary layer equation \eqref{eq:adjoint_streamfunction} is, using \eqref{eq:adjoint_variable_transformation}
\begin{equation}
        \tilde{\psi}(x,\eta) = \frac{2}{L U^2} \left( 1 - \varphi(\xi,\eta) \right) \label{eq:adjoint_momentum_final}
\end{equation}

\subsection{Analysis of the adjoint boundary conditions}
\label{sec:adjoint_analysis}

Let us analyze the adjoint solutions \eqref{eq:analytic_adjoint_drag} and \eqref{eq:adjoint_momentum_final}. It can be verified that 
they obey their respective adjoint equations (\ref{eq:linearized_adjoint_eta}) and (\ref{eq:adjoint_streamfunction}). We can also verify that \eqref{eq:analytic_adjoint_drag} obeys the adjoint boundary conditions  \eqref{eq:adjoint_eta_boundary_conditions}. The constraints on \eqref{eq:adjoint_momentum_final} are trivially satisfied as a direct consequence.

At the plate's end, $\xi=1$ and we have

\begin{equation}
    \tilde{\Psi}(L,\eta) = \frac{\kappa}{\sqrt{2L}} (\varphi(1,\eta) - 1) = 0 \label{eq:boundary_check_L}
\end{equation}
since $\varphi(1,\eta) = 1$ from (\ref{eq:D_0_definition}). 

At the wall, $\eta = 0$. Since $D_k(0) = 0$ for all $k$ we have that $\varphi(\xi,0) = 0$, $0 \le \xi < 1$, which yields:
\begin{equation}
    \tilde{\Psi}(x,0) = - \frac{\kappa}{\sqrt{2x}} \label{eq:boundary_check_wall}
\end{equation} 
as expected. 

At $\xi = 1$ there seems to be a contradiction since the wall b.c. seems to demand $\varphi(1,0) = 0$ while (\ref{eq:D_0_definition}) yields $\varphi(1,\eta) = 1$. From a physical viewpoint, this simply reflects the conflicting boundary conditions at the corner point between the outlet and the wall and is identical to the behavior of the Blasius function $F_0(y/\sqrt{x})$ as $x\to 0$ for $y\neq 0$ and $y=0$. 
%
%

Finally, vanishing of \eqref{eq:analytic_adjoint_drag} at the farfield $\eta \rightarrow \infty$ requires that $\lim_{\eta \rightarrow \infty} \varphi(\xi,\eta) = 1 $. This will be demonstrated in appendix \ref{app:properties_phi}. 

\subsection{Adjoint similarity solution}
\label{sec:adjoint_similarity}

A fundamental question arises regarding the similarity properties of the adjoint solution \eqref{eq:adjoint_momentum_final}. Self-similarity of the adjoint solution would have implications for the flow sensitivities, which would be universal up to scaling, and for the structure of the Green's function, which would likely enjoy analogous similarity properties. Since the adjoint equations are directly derived from the linearization of the self-similar Blasius equations, one might expect the dual solution to inherit this self-similarity. If the base flow is self-similar, does the adjoint solution naturally reduce to a self-similar form as well? This question is actually very hard, since the answer depends on the base flow, the objective function and the domain. We will therefore focus on the current setting (integrated viscous drag on a plate of finite length) and start by acknowledging a simplified system for which the answer is positive, and then we will move on to consider the Blasius case.

Let us consider the Oseen linearization of the boundary layer equations, which is described in more depth in Appendix A. The adjoint solution---which can be written as an eigenfunction expansion---can be put in the following similarity form:
\begin{equation}
    \tilde{\psi}(x,y) = \frac{2}{LU^2} \left[ 1 - \mathrm{erf}\left(\frac{\hat{\eta}}{\sqrt{2}}\right) \right]
\label{eq:adj_oseen_blasius}    
\end{equation}
where $\mathrm{erf}(t) = 2\pi^{-1/2} \int_0^t \exp(-z^2)dz$ is the error function and $\hat{\eta} = y\sqrt{Re_m/(2(L-x))}$ is Blasius variable measured from the end of the plate.

Turning to the adjoint Blasius equation, if a self-similar solution existed, it would be natural to assume that the similarity variable would be $\hat{\eta}$ as well. This is, in fact, the proposal of \cite{Kuhl2021}, where it is claimed that under suitable approximations, the $x$-momentum drag-based adjoint variable $\tilde{\psi}$ is self-similar with similarity variable $\hat{\eta}$ and similarity function $F_{0,\eta}$, such that:
\begin{equation}
    \tilde{\psi}(x,y) = \frac{2}{LU^2} \left[ 1 - F_{0,\eta}(\hat{\eta}) \right]
    \label{eq:kuhl_ansatz}
\end{equation}
However, \eqref{eq:kuhl_ansatz} does not obey the adjoint boundary-layer equation \eqref{eq:adjoint_streamfunction} as can be checked by direct substitution. We can further prove this fact using the eigenfunction expansion. For the Oseen equation, the eigenfunction expansion collapses to $\text{erf}(\hat{\eta}/\sqrt{2})$.
For the Blasius equation, the ansatz \eqref{eq:kuhl_ansatz} requires that:
\begin{equation}
    \sum_{k=1}^\infty \frac{D_k(\eta)}{\lambda_k - 1} \xi^{(\lambda_k - 1)/2} = F_{0,\eta} \left( \sqrt{\frac{\xi}{1-\xi}} \eta \right) \label{eq:similarity_ansatz}
\end{equation} 
To test (\ref{eq:similarity_ansatz}), let us consider the adjoint eigenfunction expansion of $F_{0,\eta}(\eta)$, i.e.
\begin{equation}
    F_{0,\eta}(\eta) = \sum_{k=1}^\infty a_k D_k(\eta) \label{eq:F_0_eta_expansion}
\end{equation} 
The coefficients $a_k$ can be obtained by integrating $F_{0,\eta}$ against the corresponding eigenfunction with the Sturm-Liouville weight $F_{0,\eta\eta}$:
\begin{equation}
    a_k = C_k \int_0^\infty F_{0,\eta\eta}(\eta) F_{0,\eta}(\eta) D_k(\eta) d\eta \label{eq:a_k_similarity}
\end{equation} 
We quote the results for the first 10 eigenfunctions obtained from the numerical integration of the Blasius and adjoint equations in Table \ref{tab:coefficients}. 
If  \eqref{eq:similarity_ansatz} held, we could set $\xi = 1/2$ and then $F_{0,\eta}$ should obey the following identity:
\begin{equation}
    F_{0,\eta}(\eta) = \sum_{k=1}^\infty \frac{D_k(\eta)}{2^{(\lambda_k - 1)/2} (\lambda_k - 1)} \label{eq:similarity_identity}
\end{equation} 
Equality of  \eqref{eq:similarity_identity} and  \eqref{eq:F_0_eta_expansion} requires that $a_k = 2^{-(\lambda_k - 1)/2} (\lambda_k - 1)^{-1}$. This is clearly not the case, as can be seen in Table \ref{tab:coefficients}. 

\begin{table}[h!]
\centering
\begin{tabular}{ccc}
\hline
$k$ & $a_k$ & $2^{-(\lambda_k-1)/2}(\lambda_k-1)^{-1}$ \\
\hline
1 & 0.78629152 & 0.707106781 \\
2 & 0.19995432 & 0.137871733 \\
3 & 0.09182893 & 0.043435906 \\
4 & 0.05407272 & 0.016064568 \\
5 & 0.03668328 & 0.006434176 \\
6 & 0.02716455 & 0.002702378 \\
7 & 0.02129176 & 0.001171264 \\
8 & 0.01734642 & 0.000518991 \\
9 & 0.01453172 & 0.000233942 \\
10 & 0.01239230 & 0.000106789 \\
\hline
\end{tabular}
\caption{Coefficients of the adjoint eigenfunction expansion of $F_{0,\eta}$}
\label{tab:coefficients}
\end{table}

The mismatch between the coefficients rules out \eqref{eq:kuhl_ansatz}. This is also confirmed by direct numerical evaluation of the semi-analytic adjoint solution in section \ref{sec:numerical_analysis}, which does not show similarity with respect to either $\hat{\eta}$ or $\eta$, nor do any computations carried out to date by the authors using a variety of different Navier-Stokes solvers. 

The fundamental reason that explains why the drag-based Blasius adjoint solution cannot collapse into a simple backward-facing similarity variable (or any other similarity variable) lies in the lack of translation symmetry. In the linear Oseen approximation, the constant background velocity $U$ makes the problem translation-invariant in $x$; consequently, the adjoint solution depends solely on the distance from the trailing edge ($L-x$) and a similarity variable like $\hat{\eta} \sim y/\sqrt{L - x}$ can be expected. The Blasius solution breaks this picture. The leading edge at $x = 0$ is a privileged point, and the boundary layer evolves continuously downstream. Since the adjoint equations are built around the linearization of this base state, their coefficients explicitly depend on the distance from this origin. The adjoint equations thus have two conflicting longitudinal scales, $x$ and $L- x$. A single similarity variable cannot simultaneously reconcile the forward-evolving flow and the backward-propagating adjoint field, precluding a self-similar collapse.

\section{Numerical analysis}
\label{sec:numerical_analysis}

Here we focus on explicitly building the adjoint solution \eqref{eq:adjoint_momentum_final} in order to verify its properties and to compare it to numerical adjoint solutions. 
The evaluation of these expressions requires the construction of the adjoint eigenfunctions and their corresponding eigenvalues and norms. 

%
%

In order to build the eigenfunctions we proceed as follows.
Eigenvalues up to $n=20$ are taken from Luchini \cite{Luchini1996} and Wilks and Bramley \cite{WilksBramley1977}, while for $n>20$, Brown's asymptotic relation (\ref{eq:brown_formula}) is used. 
These values are used to compute the adjoint eigenfunctions by numerical integration of either equations (\ref{eq:eigenvalue_problem}) or (\ref{eq:adjoint_eigenvalue_problem}). The normalizing constants $C_n$ are obtained by numerical evaluation of integrals (\ref{eq:orthogonality}) or (\ref{eq:adjoint_orthogonality}). Their values are found to be reasonably well correlated by the equation 
\begin{equation}
    C_n \sim 1.443 n^{-0.672} \label{eq:norm_correlation}
\end{equation} 
(or $C_n \sim 2.2441 \lambda_n^{-0.669}$ in terms of the eigenvalues), which compares favorably with Kemp's approximate formula $C_n = 1.37 n^{-2/3}$ \cite{Kemp1970}

%
%

\begin{figure}[H]
    \centering
    \includegraphics[width=0.48\textwidth]{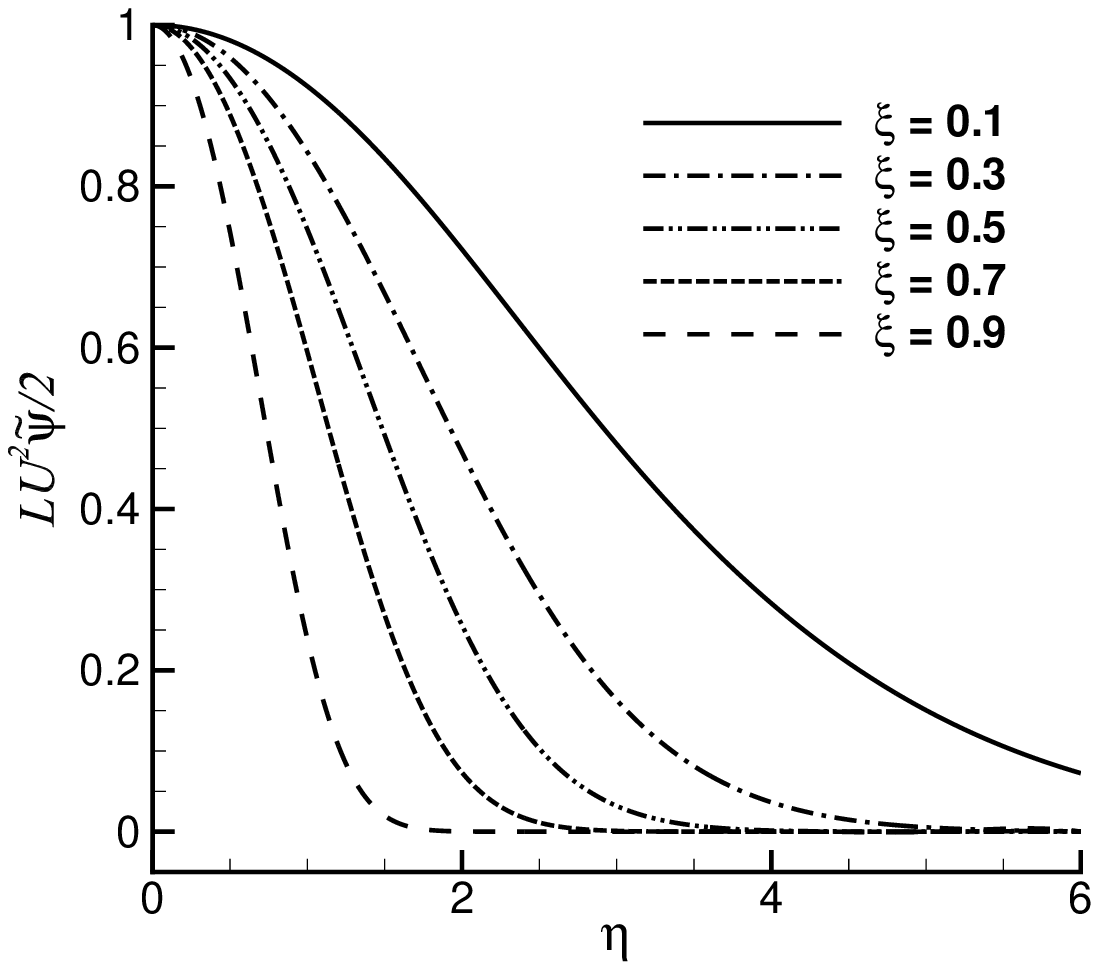}
    \includegraphics[width=0.48\textwidth]{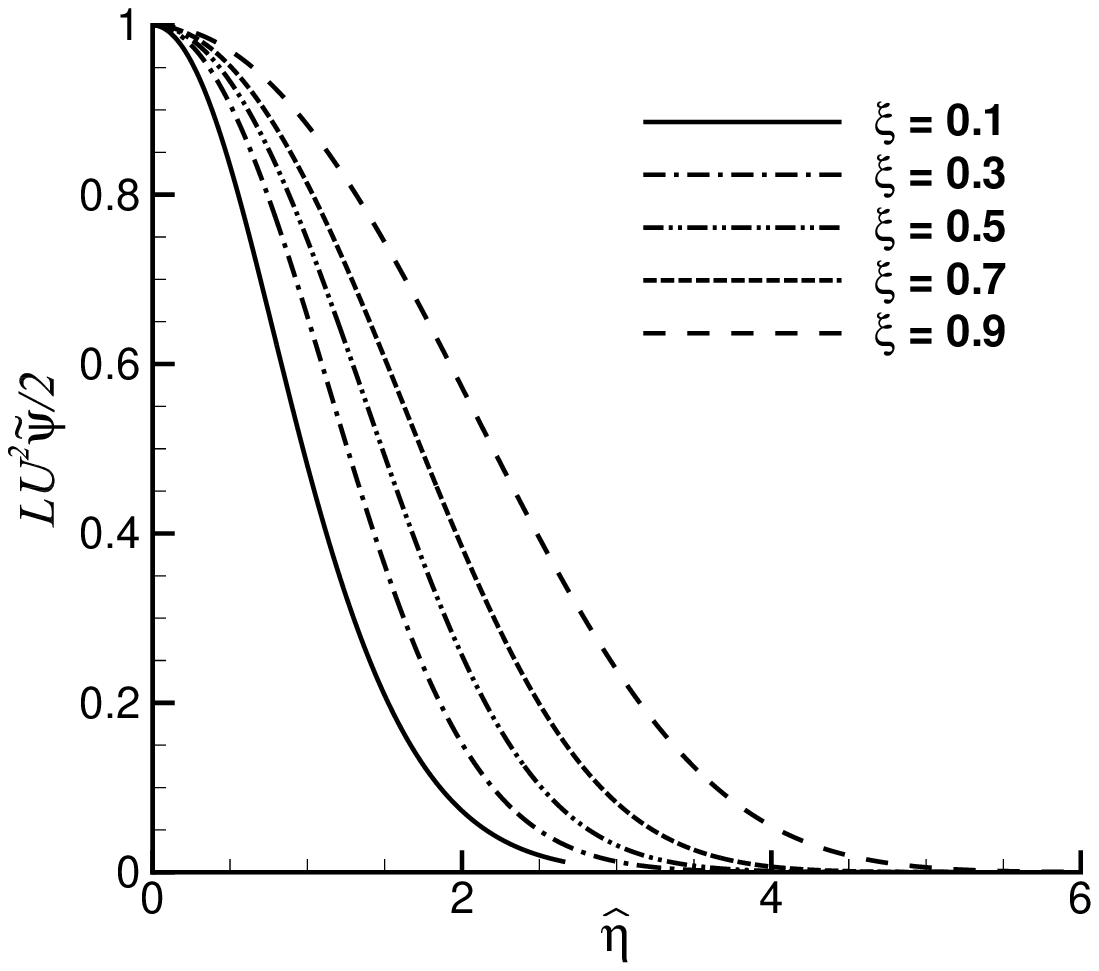}
    \caption{Approximate value of the adjoint solution $L U^2 \tilde{\psi}/2$ computed with 2800 eigenfunctions plotted against $\eta$ (left) and $\hat{\eta}$ (right).}
    \label{fig:fig1}
\end{figure}

Using these computed eigenvalues and norms, we construct the sample solutions. Figure \ref{fig:fig1} shows the adjoint solution (\ref{eq:adjoint_momentum_final}) computed with $N_{eigen} = 2800$ eigenfunctions. The solution is plotted against $\eta$ and $\hat{\eta}$. It is clear that there is no similarity in either variable, even though the solution only depends on $x$ and $y$ through the dimensionless variables $\xi$ and $\eta$. Notice also that $LU^2\tilde{\psi}/2$ seems to be bounded between 0 and 1, which means that the function $\varphi(\xi,\eta)$ defined in  (\ref{eq:phi_definition}) is also bounded between 0 and 1. We will show this to be the case in appendix \ref{app:properties_phi}. 

\begin{figure}[H]
    \centering
    \includegraphics[width=0.7\textwidth]{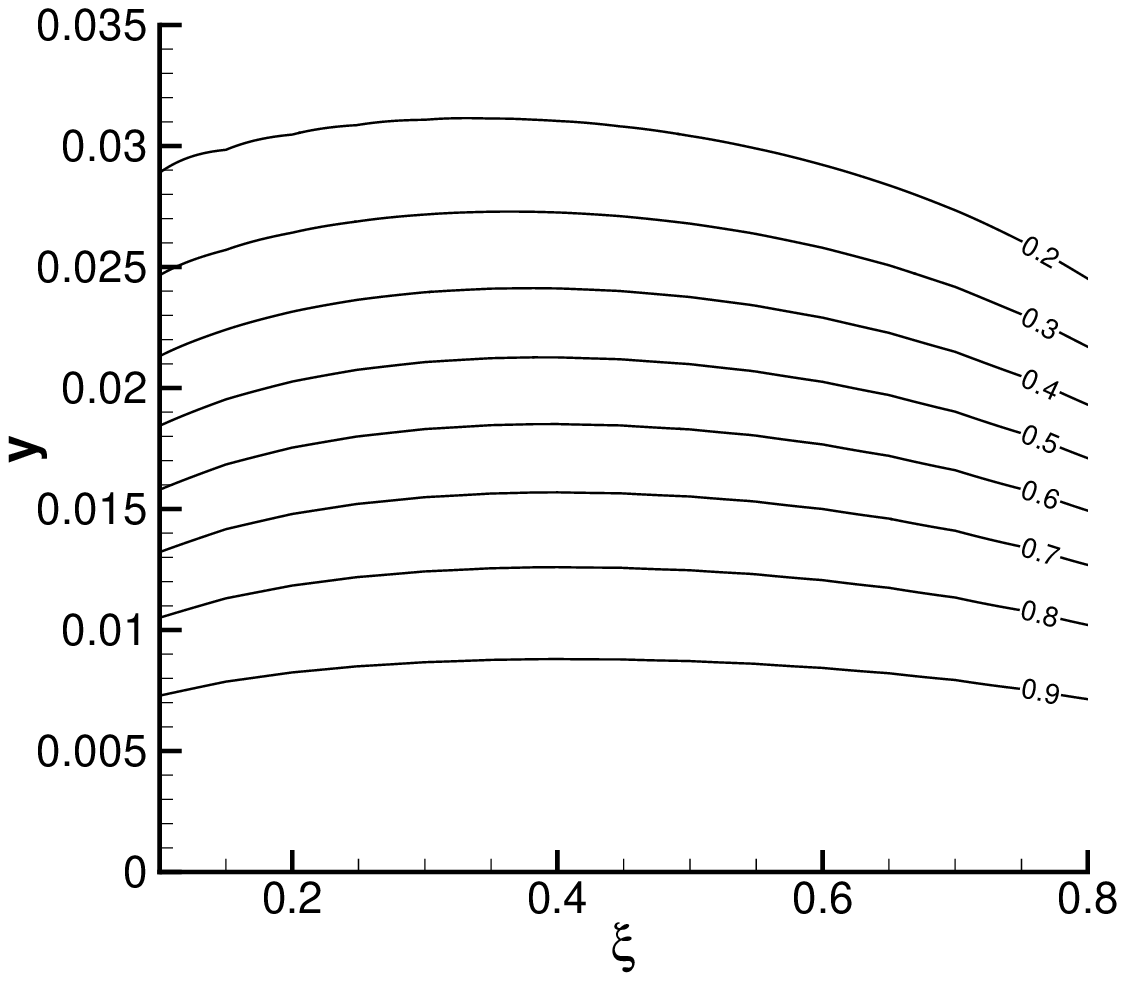}
    \caption{Contour levels of the adjoint solution $L U^2 \tilde{\psi}/2$ computed with 2800 eigenfunctions.}
    \label{fig:fig1ymedio}
\end{figure}

The spatial structure of the solution is actually more involved than expected. Instead of a simple boundary layer structure emanating from the trailing edge, the adjoint contours plotted in figure \ref{fig:fig1ymedio} show a parabolic-like shape peaking roughly around 1/3 of the plate's length and decaying towards the leading and trailing edges. This behavior is likely a direct physical manifestation of the two streamwide length scales $x$ and $L-x$ that we discussed in section \ref{sec:adjoint_similarity}.  

\begin{figure}[H]
    \centering
    \includegraphics[width=0.48\textwidth]{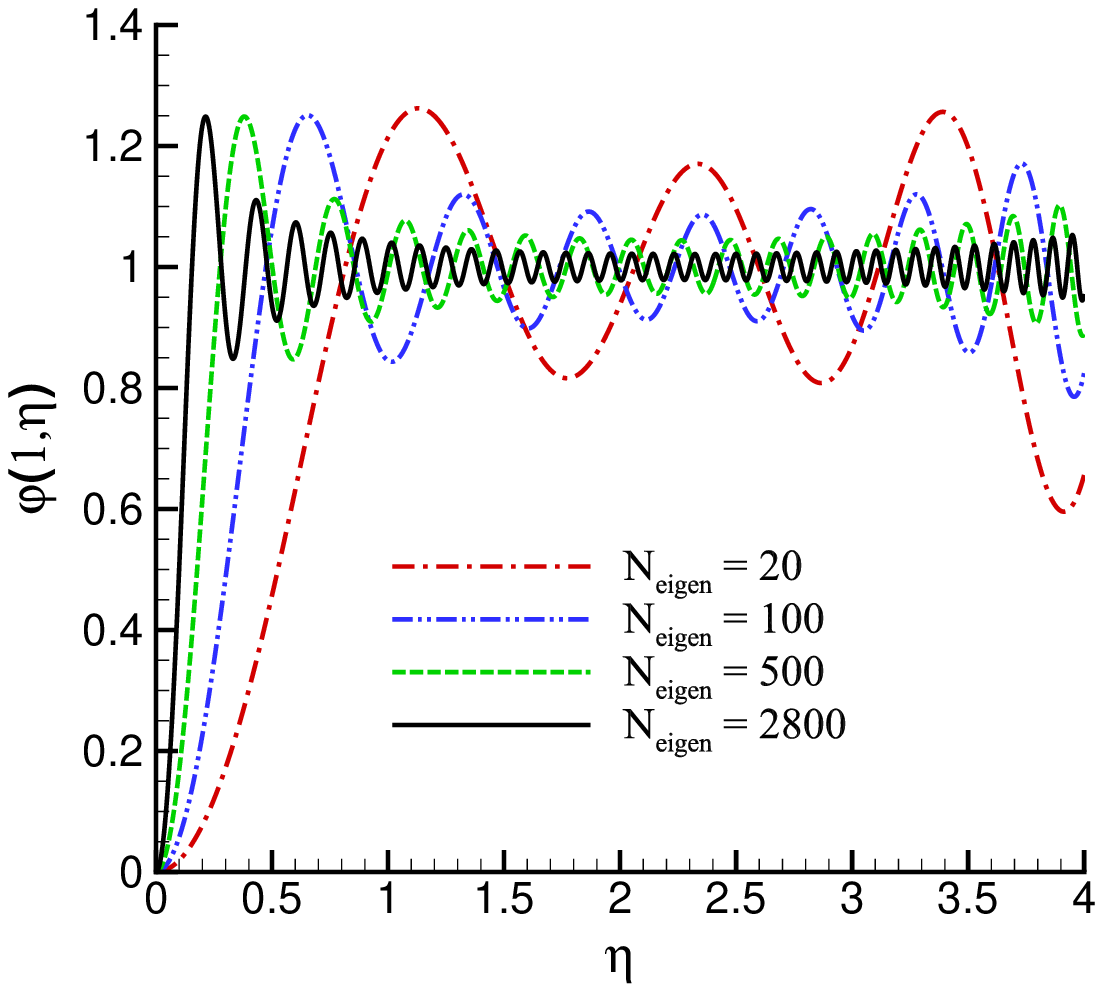}
    \includegraphics[width=0.48\textwidth]{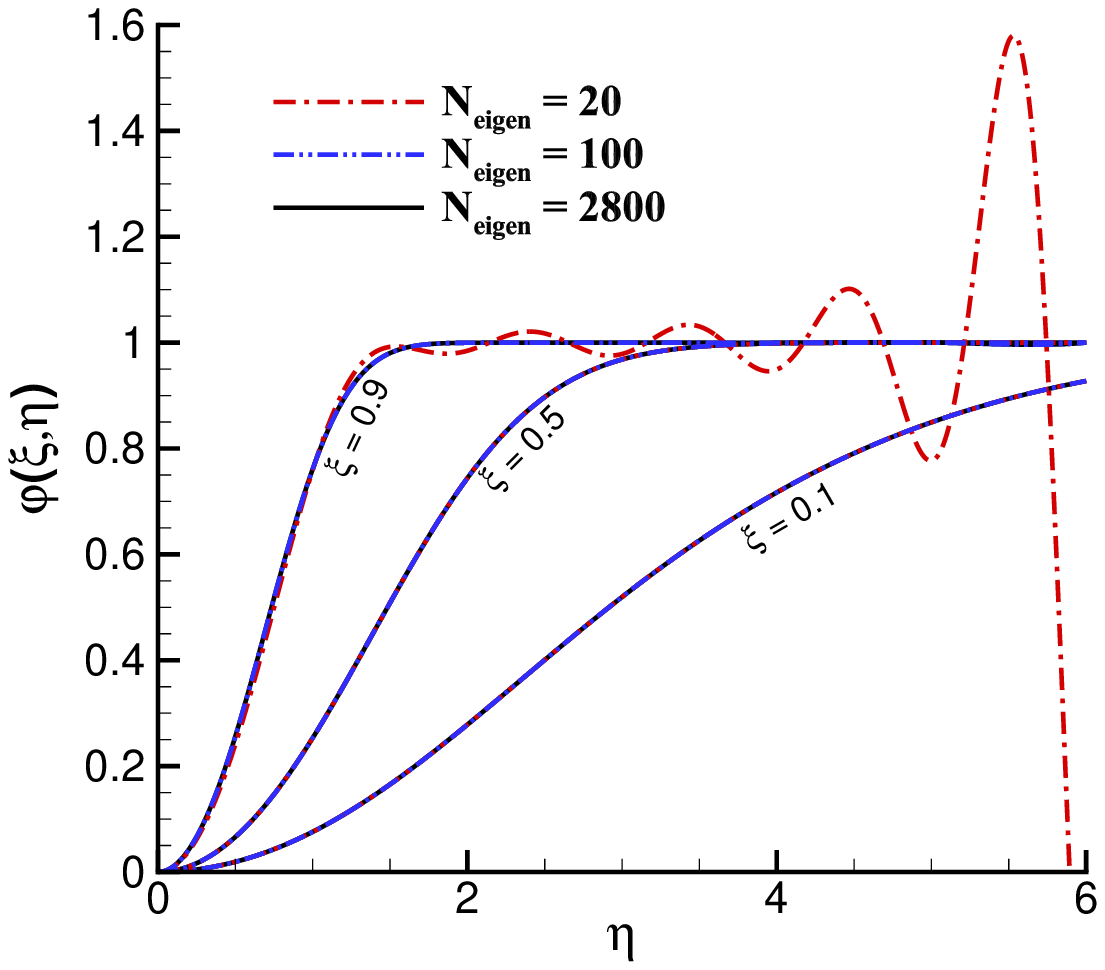}
    \caption{Analysis of the convergence of the eigenfunction expansion. Approximate value of the function $\varphi(\xi,\eta)$ for different number of eigenfunctions and several positions along the plate. Left: $\xi=x/L=1$. Right: $\xi=0.1, 0.5$ and $0.9$.}
    \label{fig:fig2}
\end{figure}

It is interesting to ask about the rate of convergence of the eigenfunction expansion. 
In \cite{LibbyFox1963}, Libby and Fox pointed out the slow convergence of the expansion in various perturbation problems.
Figure \ref{fig:fig2} focuses on this issue, plotting the function $\varphi(\xi,\eta)$ evaluated with different number $N_{eigen}$ of eigenfunctions. 
The convergence is rapid except near the end of the plate, even though for $\xi=0.9$ the solution with 100 eigenfunctions is virtually indistinguishable from the solution with 2800 eigenfunctions.
On the other hand, for $\xi=1$ the convergence is quite slow, with strong oscillations typical of the Gibbs phenomenon even with 2800 eigenfunctions, which however show a convergence trend towards the desired value of 1. 

\begin{figure}[htbp]
    \centering
    \includegraphics[width=0.7\textwidth]{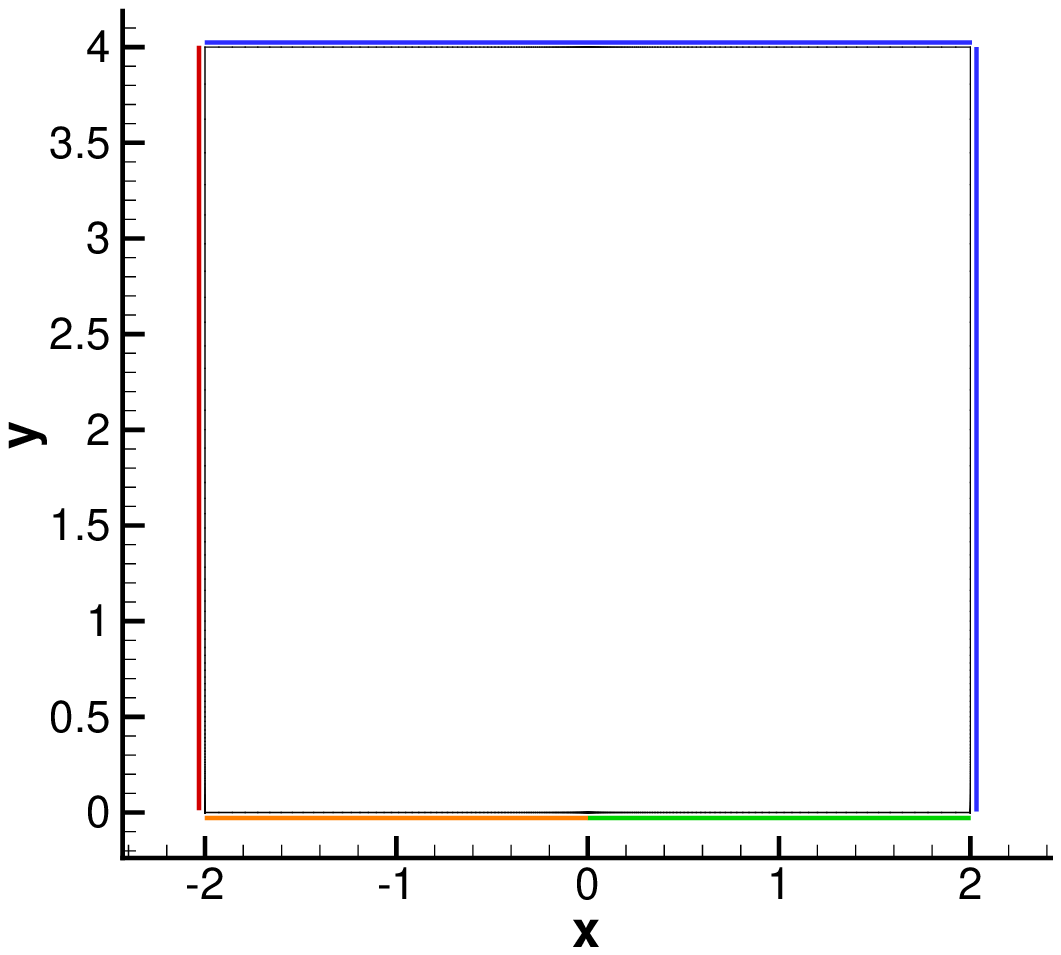}
    \caption{Sketch of the computational domain with boundary conditions: inlet (red), outlet (blue), symmetry (orange), wall (green).}
    \label{fig:nishi_mesh}
\end{figure}

In order to verify the proposed semi-analytic adjoint solution, we compare the theoretical profiles against numerical data obtained by discretizing the adjoint incompressible boundary layer and full Navier-Stokes equations, respectively. Both solvers have been assembled from H. Nishikawa's edu solver \cite{Nishikawa2020}, using a cell-centered finite-volume discretization with Roe's flux and the alpha-damping viscous flux \cite{Nishikawa2010}. In both cases, the simulation has been performed at $Re_m=10000$ on a $272 \times 192$ mesh representing a $2L \times 2L$ computational domain with total plate length $L=2$ and an equal buffer zone upstream of the plate. 
For the flow solvers, total pressure and velocity direction are imposed at the inlet (left) boundary, while static pressure is extrapolated from the interior. 
At the outer (top and right) boundaries, static pressure is prescribed and velocity is extrapolated from the interior.
Symmetry conditions are used along the part of the bottom boundary upstream of the plate. 
For the adjoint solvers, dual inlet and outlet conditions are imposed as explained in \cite{LozanoPonsin2012}. A scheme of the computational domain and the boundary conditions is shown in figure \ref{fig:nishi_mesh}.

\begin{figure}[H]
    \centering
    \includegraphics[width=0.48\textwidth]{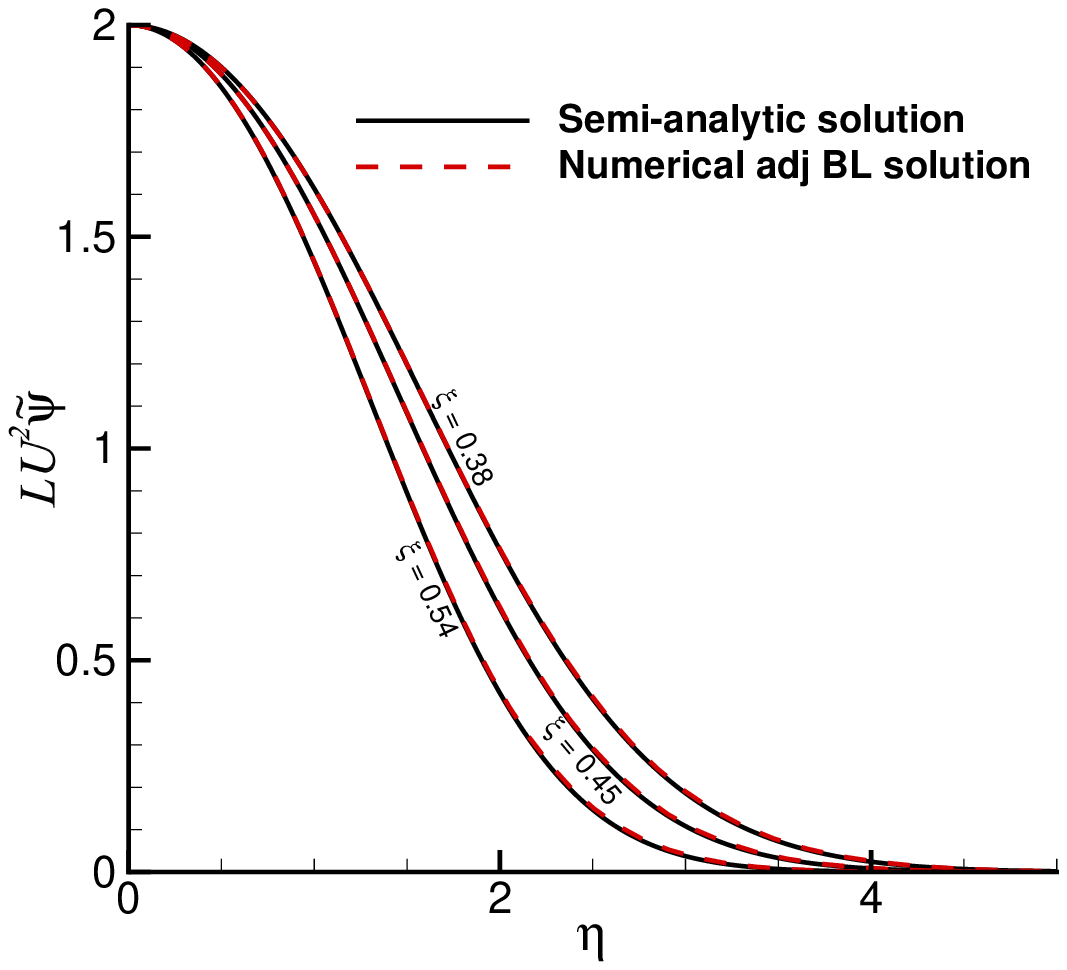}
    \includegraphics[width=0.48\textwidth]{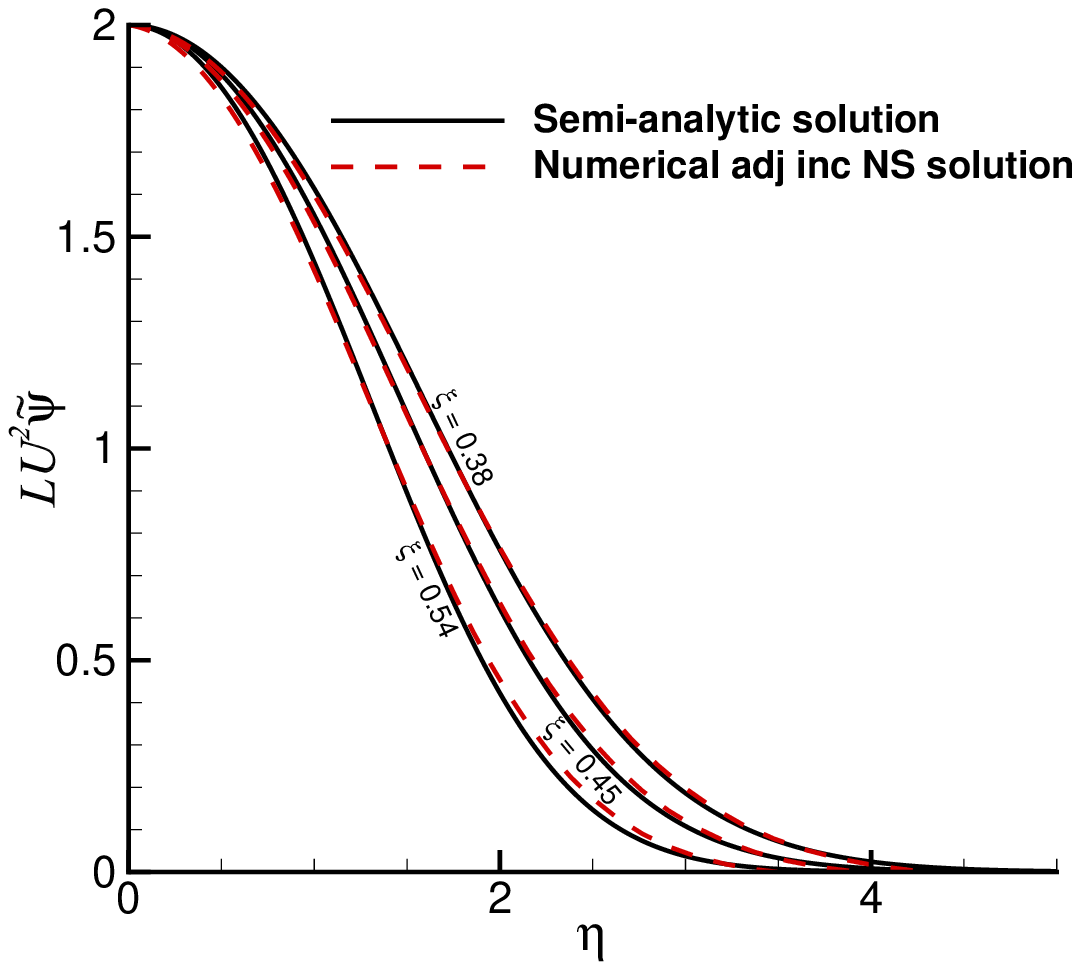}   
    \caption{Semi-analytic adjoint solution vs numerical incompressible adjoint solutions. Left: BL data. Right: full NS data.}
    \label{fig:fig3}
\end{figure}

Figure \ref{fig:fig3} shows the comparison between the semi-analytic solution and the numerical solutions obtained with the BL (left) and NS (right) solvers, respectively. We see that the theoretical solution shows excelent agreement with the adjoint boundary layer results, while also remaining highly robust when compared against the full NS system.
 
%
%

\begin{figure}[htbp]
    \centering
    \includegraphics[width=0.48\textwidth, trim=1.1cm 0.5cm 2.4cm 3.9cm, clip]{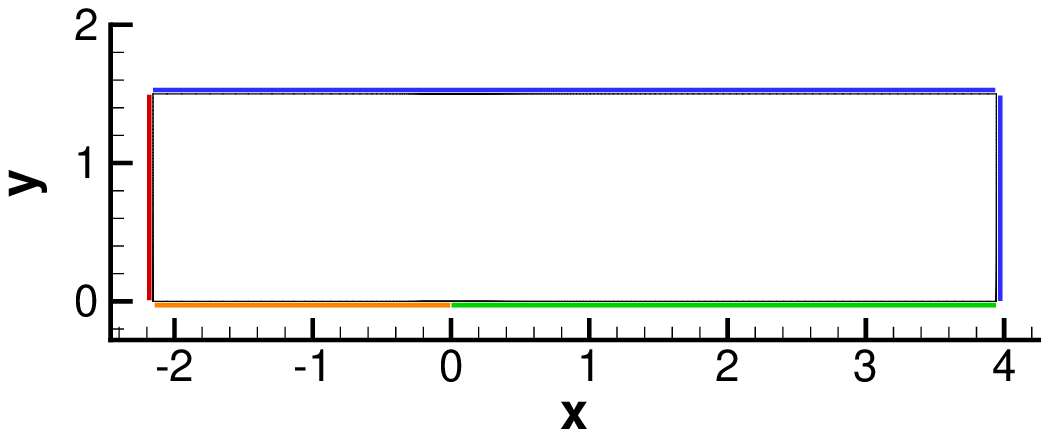}
    \hfill
    \includegraphics[width=0.48\textwidth]{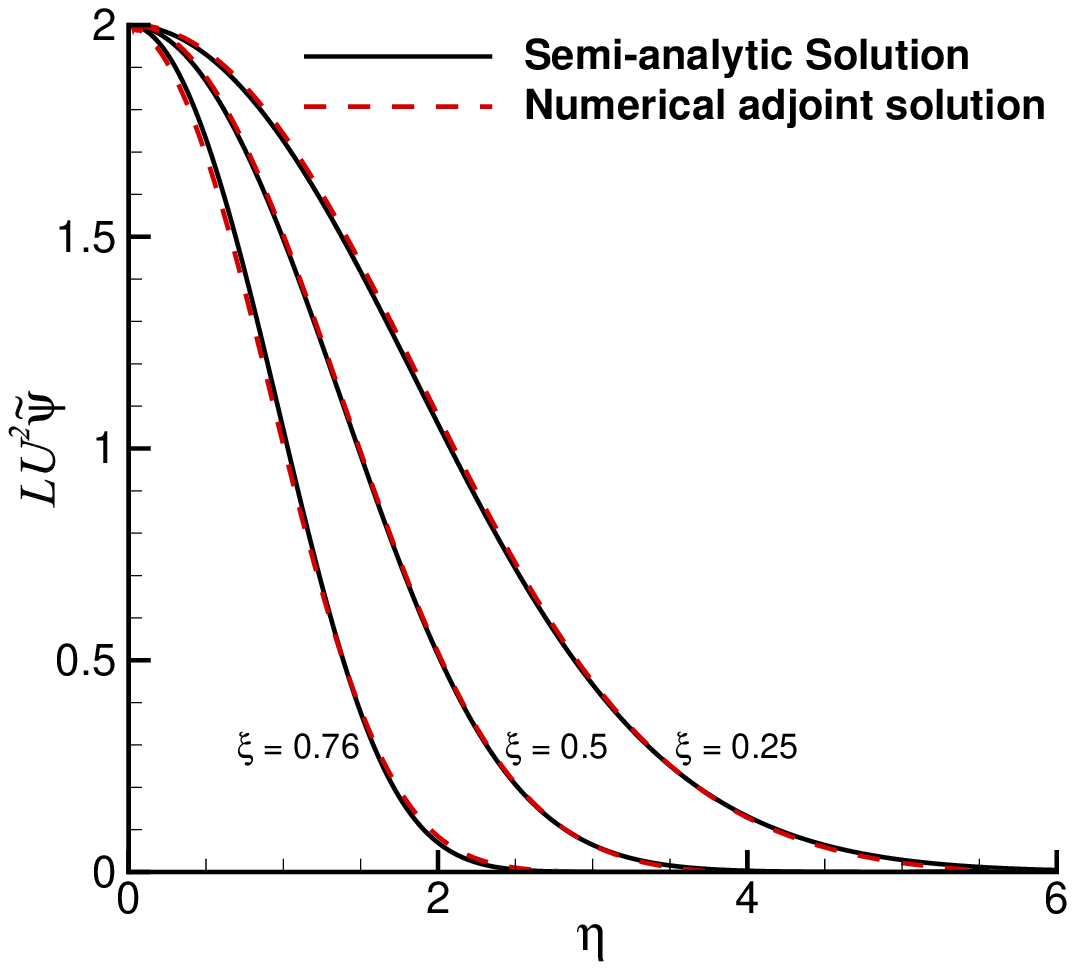}
    \caption{Compressible flat plate computation. Left: Detail of computational domain with boundary conditions: inlet (red), outlet (blue), symmetry (orange), wall (green). Right: comparison of numerical and semi-analytic adjoint solutions.}
    \label{fig:fig4}
\end{figure}

Finally, figure \ref{fig:fig4} shows a comparison against numerical results obtained with a compressible Navier-Stokes solver (INTA's finite-volume NENS2D solver described and validated in \cite{Castro2007}) using a central scheme with Jameson-Schmidt-Turkel (JST)-type scalar artificial dissipation and Weiss \cite{Weiss1997} node-gradient-based viscous fluxes. 
The simulation was carried out at $Re_m=15000$ and $M=0.2$ on a $437 \times 313$ mesh representing a $1.55L \times 0.375L$ domain with total plate length $L=4$. A buffer zone of length $0.55L$ with symmetry boundary condition along the bottom boundary was set upstream of the plate.
Total pressure and temperature and flow direction are imposed at the inlet (left) boundary, while Mach number is extrapolated from the interior.
At the outlet (top and right) boundary, static pressure is imposed and density and velocity are extrapolated from the interior. 
For the adjoint solver, the corresponding dual boundary conditions are used. Details of the computational domain are shown in figure \ref{fig:fig4}.  While this is a completely different problem, the agreement of the semi-analytic solution with this low-Mach compressible computation is nevertheless quite good, demonstrating that the proposed solution is an acceptable tool for validation of compressible adjoint NS solvers as well.

%
%

\section{Applications}
\label{sec:applications}

In this section, the adjoint solution is applied to analyze the Adjoint Transport Convection (ATC) term and to evaluate flow sensitivities for shape design, initial-value perturbations and active flow control.

\subsection{Analysis of the adjoint transport convection term (ATC)}
\label{sec:atc}

The Adjoint Transport Convection (ATC) term is a specific cross-coupling term that appears in the continuous adjoint formulation of the Navier-Stokes momentum equations. 
For the full incompressible Navier-Stokes equations, this term has the form:
\begin{equation}
    ATC = \begin{pmatrix} u\tilde{\psi}_x + v\tilde{\zeta}_x \\ u\tilde{\psi}_y + v\tilde{\zeta}_y \end{pmatrix} \label{eq:atc_definition}
\end{equation} 
where $(\tilde{\psi}, \tilde{\zeta})$ are the $x$- and $y$-momentum adjoint variables, linked by the adjoint continuity equation $\tilde{\psi}_x + \tilde{\zeta}_y = 0$. 
In  \eqref{eq:atc_definition}, the two rows correspond to the $x$ and $y$-momentum adjoint equations, respectively. 

The ATC term is often a source of numerical instabilities in adjoint solvers on typical industrial meshes \cite{SotoLohner2004, Othmer2014}. 
Various numerical strategies, including damping or neglecting the ATC, have been used at the cost of trading robustness for accuracy while preserving the qualitative correctness of the sensitivities. 
Neglecting this term causes the adjoint problem to have the same convective structure as the primal flow so that the same numerical technique can be used to solve both the flow and the adjoint equations. 

The ATC was discussed in \cite{Kuhl2021}, where it was concluded that, for the Blasius problem, the ATC actually disappears, and that its vanishing was linked to the adjoint solution being self-similar with a similarity profile equal to that of the primal flow. As argued in Section \ref{sec:adjoint_similarity}, this assumption does not hold.   

%

Since we have the semi-analytic solution at our disposal, we can evaluate the ATC directly to quantify its true magnitude. While our solution provides the $x$-momentum adjoint variable $\tilde{\psi}$, the $y$-momentum adjoint variable $\tilde{\zeta}$ is constrained by the continuity equation $\tilde{\psi}_x + \tilde{\zeta}_y = 0$. This means that the $y$-momentum ATC can be estimated as $u\tilde{\psi}_y - v\tilde{\psi}_x$, while the $x$-momentum ATC $u\tilde{\psi}_x + v\tilde{\zeta}_x$ still contains the unknown contribution $\tilde{\zeta}_x$. Applying standard boundary-layer scaling, we find that $\tilde{\psi} \sim (L/\delta)\tilde{\zeta}$, where $\delta$ is the boundary layer thickness. Therefore, the terms involving $v\tilde{\zeta}_x$ and $v\tilde{\zeta}_y$ are smaller by a factor of $(\delta/L)^2$ relative to the primary terms, making them physically negligible at high Reynolds numbers.

\begin{figure}[H]
    \centering
    \includegraphics[width=0.48\textwidth]{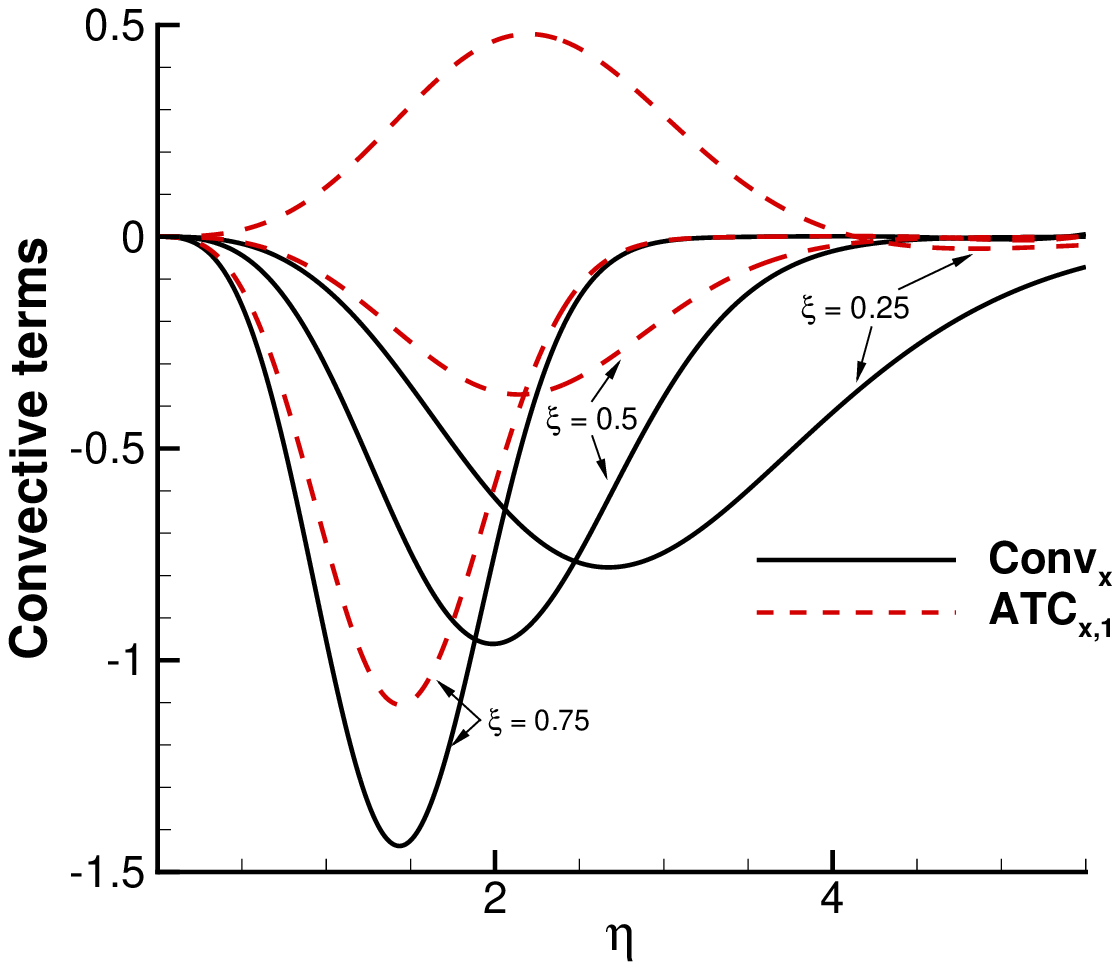}
    \includegraphics[width=0.48\textwidth]{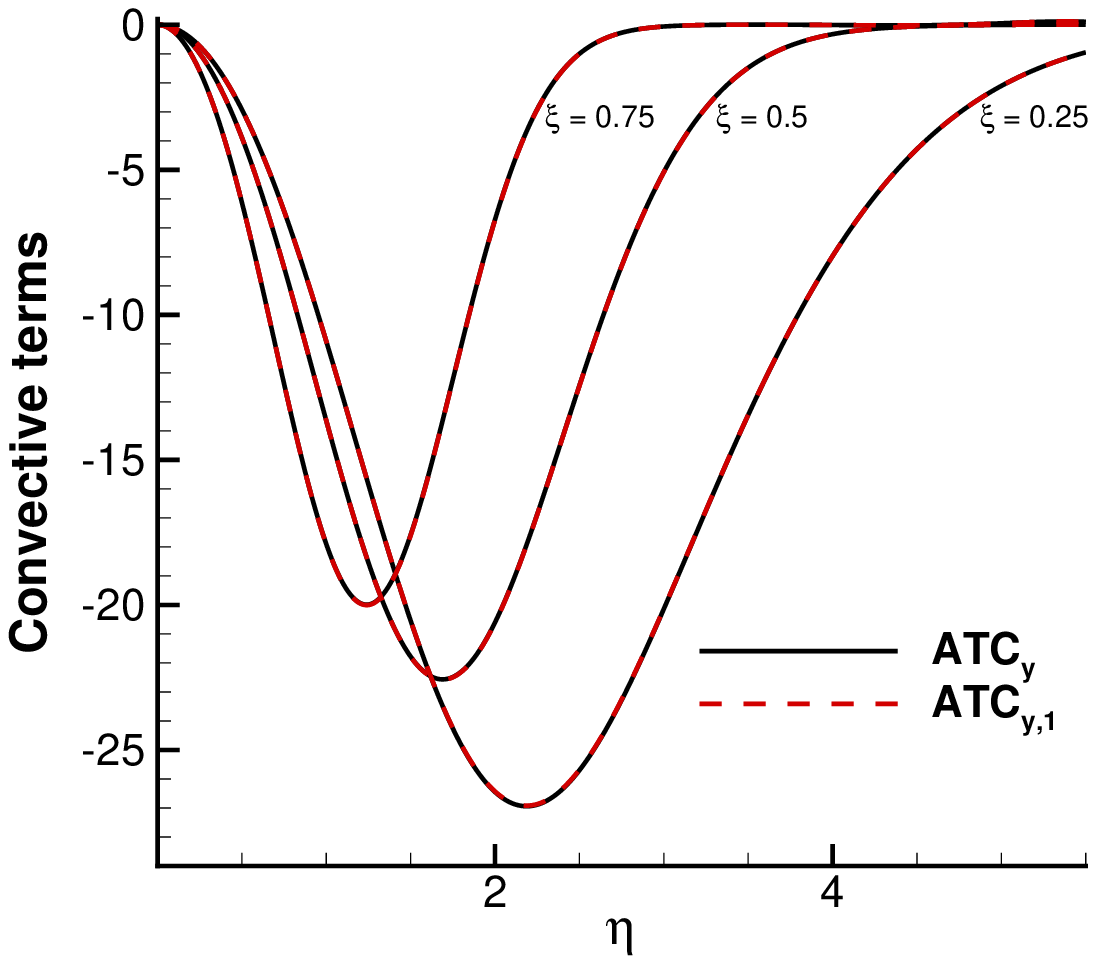}
    \caption{Left: ATC and convective term for the $x$-momentum adjoint equation. Right: ATC for the $y$-momentum adjoint equation.}
    \label{fig:fig5}
\end{figure}

This scaling is explicitly confirmed in figure \ref{fig:fig5}, which compares the ATC against the standard convective terms for $Re_m = 10000$. The left panel demonstrates that the dominant $x$-momentum ATC component, $ATC_{x,1}=u\tilde{\psi}_x$, is of the same order of magnitude as the full standard convective term $Conv_x=u\tilde{\psi}_x + v\tilde{\psi}_y$. According to the BL scaling argument, the missing $v\tilde{\zeta}_x$ contribution is 10000 times smaller than $u\tilde{\psi}_x$.

The right panel, on the other hand, compares the full $ATC_y = u\tilde{\psi}_y + v\tilde{\zeta}_y = u\tilde{\psi}_y - v\tilde{\psi}_x$ with its largest contributing part $ATC_{y,1} = u\tilde{\psi}_y$. 
The term depending on the $y$-momentum variable, $v\tilde{\zeta}_y$, is again 10000 times smaller than $u\tilde{\psi}_y$, so both lines are virtually indistinguishable. 
The convective term for the $y$-momentum equation, $Conv_y = u\tilde{\zeta}_x + v\tilde{\zeta}_y$, is not shown here, even though it can be seen to be $\mathcal{O}(\delta/L)^2 \approx 1/10000$ smaller than the ATC for $Re_m=10000$ by the same scaling argument. 

 The preceeding analysis shows that even in the simplified context of the Blasius 
boundary layer, the ATC term computed with the semi-analytic adjoint solution is comparable in size to the regular adjoint convective term and cannot be neglected. Similar analysis carried out with the numerical solution, that we do not present here, show analogous conclusions. Therefore, our findings suggest that the omission of the ATC in numerical solvers should be done with care, as it signicantly alters the adjoint momentum balance.

\subsection{Shape sensitivity of the integrated friction drag coefficient}
\label{sec:shape_sens}

The semi-analytic adjoint solution can be used to evaluate the sensitivity of the total wall drag coefficient to infinitesimal changes in the shape of the plate. It must be borne in mind that introducing geometric deformations to the plate's geometry can induce instabilities and trigger transition to turbulence. The present analysis is thus based in the assumption that the geometric deformations are smooth and sufficiently small so that the perturbed flow is laminar and deviates only slightly from the Blasius solution.  

Under these assumptions, and provided  \eqref{eq:adjoint_streamfunction} and  \eqref{eq:adjoint_boundary_conditions} hold, the shape sensitivity derivative can be computed from  \eqref{eq:linearized_lagrangian} as (see also \cite{Castro2007, SotoLohner2004}):
\begin{equation}
    \delta C_D(L) = -\int_0^L \nu k (\tilde{\psi} \psi_{yy})_{y=0} \delta x_n dx - \int_0^L \nu (\tilde{\psi} \delta x_n \psi_{yyy})_{y=0} dx + \int_0^L dx \nu \delta x_n \tilde{\psi}_y \psi_{yy} \label{eq:shape_sensitivity_integral}
\end{equation} 
where $k$ is the curvature of the unperturbed geometry \cite{Castro2007} and $\delta x_n$ is the shape deformation in the local normal direction. 
The first two terms in  \eqref{eq:shape_sensitivity_integral} come from the geometric variation of the cost function (with the adjoint wall b.c.  \eqref{eq:adjoint_boundary_conditions} used to replace $\tilde{\kappa} = \nu\tilde{\psi}_{wall}$ as in \cite{SotoLohner2004}), while the last one comes from the integration by parts required to derive the adjoint equation \eqref{eq:adjoint_streamfunction} and is proportional to the product of the flow and adjoint shears \cite{Kuhl2021}. 

The leftmost term on the right-hand side (RHS) of (\ref{eq:shape_sensitivity_integral}) is zero for the flat plate since the curvature is zero. The middle term is also zero since $\psi_{yyy}$  is proportional to $F_{0,\eta\eta\eta}$, which vanishes at the wall. Finally, the rightmost term also vanishes because ${\tilde{\psi}}_y(x,0) = 0$ for all $x < L$ (as follows from \eqref{eq:adjoint_momentum_final}). Consequently, at this formal level of approximation and within the steady 2D boundary layer framework, the sensitivity of the integrated friction drag to infinitesimal changes in the shape of the plate is zero.

\subsection{Initial-value problem}

Libby and Fox analysis \cite{LibbyFox1963} originated from the study of the initial value problem of the laminar boundary layer, which investigated the boundary layer development downstream of an arbitrarily specified initial velocity profile that deviates only slightly from the Blasius solution. 
One of the cases discussed by Libby and Fox  concerned the flow over a two-dimensional, permeable plate followed at $x=x_i$ by an impermeable surface. The perturbation to the skin friction along the impermeable surface was found to be given, to first order, by an expansion of the form $\delta c_f = \sqrt{2\nu/(Ux)} \sum_{k=1}^\infty A_k (x/x_i)^{-\lambda_k/2}$, where $A_k$ represents the weight of the mode $N_k$ in the eigenfunction expansion of the perturbation $\delta F$ at $x=x_i$. Integrating this over a section of length $L$ yields the total drag perturbation:
\begin{equation}
    \delta C_D = \frac{2}{L} \sqrt{\frac{2\nu}{U}} \sum_{k=1}^\infty \frac{A_k x_i^{\lambda_k/2}}{(\lambda_k - 1)} \left( x_i^{(1-\lambda_k)/2} - (x_i+L)^{(1-\lambda_k)/2} \right) \label{eq:drag_perturbation_integral}
\end{equation}

This classical problem can also be approached from the adjoint viewpoint. Considering a domain extending from $x=x_i$ to $x=x_i+L$, the perturbation to the integrated drag resulting from an initial perturbation $\delta\psi$ at $x=x_i$ can be computed directly from the boundary terms of the linearized Lagrangian \eqref{eq:linearized_lagrangian} as:
\begin{equation}
    \delta C_D = \int_0^\infty dy \Big[ \tilde{\psi} \psi_{y} \delta\psi_y \Big]_{x_i} - \int_0^\infty dy \Big[ \tilde{\psi} \psi_{yy} \delta\psi \Big]_{x_i}  \label{eq:drag_perturbation_adjoint2}
\end{equation}
Setting $\psi = \sqrt{2 Re_m \nu^2 x} F_0(\eta)$ and $\delta\psi = \sqrt{2 Re_m \nu^2 x} \delta F(x,\eta)$, inserting the analytic adjoint solution \eqref{eq:adjoint_momentum_final}, and changing the integration variable to $\eta$ yields exactly the Libby-Fox result \eqref{eq:drag_perturbation_integral}. This result confirms that the semi-analytic adjoint solution correctly reproduces the sensitivities derived via perturbation theory.

\subsection{Sensitivity to blowing and suction}
\label{sec:blowing_suction}

The adjoint solution can also be used to evaluate the sensitivity of the drag coefficient to active flow control via steady wall-normal suction or blowing  \cite{Vasile2013}. Let us examine this issue from the framework presented in this paper. 
%
%
%
As with the shape sensitivity, we must note that introducing a non-zero normal velocity $v_w(x)$ at the wall can alter the stability characteristics of the boundary layer. The following analysis formally assumes that the imposed $v_w(x)$ is continuous and sufficiently small such that the flow remains close to the Blasius solution.


Introducing a perturbation streamfunction $\delta\psi$ such that $(\delta\psi_y, -\delta\psi_x) = (0, v_w(x))$ at the wall, the drag sensitivity is given by the following boundary terms in the linearized Lagrangian \eqref{eq:linearized_lagrangian}:
\begin{equation}
    \delta C_D = -\int_0^L dx \nu \Big[ \tilde{\psi}_{yy} \delta\psi \Big]_{y=0} \label{eq:drag_blowing_simplified}
\end{equation}

%
%
%

Using  \eqref{eq:adjoint_momentum_final} to evaluate the adjoint wall shear explicitly, we obtain from  \eqref{eq:drag_blowing_simplified}:
\begin{equation}
    \delta C_D = -\frac{1}{LU} \int_0^L \frac{dx}{x} \sum_{k=1}^\infty \frac{(x/L)^{(\lambda_k - 1)/2}}{C_k (\lambda_k - 1)} \int_0^x v_w(x') dx' \label{eq:drag_blowing_final}
\end{equation}

We can validate this expression by considering two test cases analyzed by Libby and Fox:
\vspace{0.5em}

\noindent \textbf{Case 1: Similar suction/blowing profile.} We assume $v_w = v_w^* (U\nu/x)^{1/2}$, where $v_w^*$ is the non-dimensional suction-blowing parameter, ensuring the perturbed boundary layer remains self-similar \cite{White1991}. Substituting this continuous distribution into \eqref{eq:drag_blowing_final} and integrating yields:
\begin{equation}
    \delta C_D = -\frac{4 v_w^*}{Re_L^{1/2}} \sum_{k=1}^\infty \frac{1}{C_k \lambda_k (\lambda_k - 1)} \approx -2.892 \frac{v_w^*}{Re_L^{1/2}} \label{eq:drag_case1_numeric}
\end{equation}
This agrees nicely with the first-order perturbation solution obtained by Libby and Fox \cite{LibbyFox1963, Libby1970}, $\delta C_D = -4 v_w^* \hat{F}_{\eta\eta}(0) / Re_L^{1/2}$ with $\hat{F}_{\eta\eta}(0) = 0.723483$, where the function $\hat{F}(\eta)$ is the solution to the equation $\hat{F}_{\eta\eta\eta} + F_0 \hat{F}_{\eta\eta} + F_{0,\eta\eta} \hat{F} = 0$; $\hat{F}(0) = 1, \hat{F}_\eta(0) = \hat{F}_\eta(\infty) = 0$. Comparison with (\ref{eq:drag_case1_numeric}) yields the exact numerical identity 
\begin{equation}
    \sum_{k=1}^\infty \frac{1}{C_k \lambda_k (\lambda_k - 1)} = \hat{F}_{\eta\eta}(0) = 0.723483 \label{eq:identity_F_hat}
\end{equation} 

\vspace{0.5em}
\noindent \textbf{Case 2: Uniform continuous suction.} We assume $v_w = \text{const} < 0$ along the entire plate. In this case, \eqref{eq:drag_blowing_final} integrates to:
\begin{equation}
    \delta C_D = -\frac{2 v_w}{U} \sum_{k=1}^\infty \frac{1}{C_k (\lambda_k - 1)(\lambda_k + 1)} \approx -1.224 \frac{v_w}{U} \label{eq:drag_case2_numeric}
\end{equation}
Again, this result is in excellent agreement with the Libby-Fox first-order solution \cite{LibbyFox1963} 
    $\delta C_D = -v_w N_{3,1,\eta\eta}(0)/U $, with $N_{3,1,\eta\eta}(0) = 1.2243$, where $N_{3,1}(\eta)$ is the solution to the eigenvalue problem 
$
    N_{3,\eta\eta\eta} + F_0 N_{3,\eta\eta} - \lambda F_{0,\eta} N_{3,\eta} + (1 + \lambda) F_{0,\eta\eta} N_3 = 0$; 
    $N_3(0) = 1, N_{3,\eta}(0) = N_{3,\eta}(\infty) = 0$ for $\lambda = 1$.
Comparison with (\ref{eq:drag_case2_numeric}) yields a second explicit identity for the norms and eigenvalues: 
\begin{equation}
    \sum_{k=1}^\infty \frac{1}{C_k (\lambda_k - 1)(\lambda_k + 1)} = N_{3,1,\eta\eta}(0) / 2 = 0.61215 \label{eq:identity_N_31}
\end{equation}

\section{Extension to Falkner-Skan flows}
\label{sec:falkner_skan}

In \cite{ChenLibby1968, ChenLibby1968_b, Kotorynski1969}, the analysis of perturbations to boundary layers with uniform external streams was extended to flows described by the Falkner-Skan equation. This section briefly outlines how our adjoint formulation can be extended to these non-zero pressure gradient flows. It is important to note upfront that while the resulting eigenfunction expansion is a solution to the adjoint equations, it exhibits severe convergence issues. We think that documenting these problems is relevant to establish the practical limitations of the semi-analytic approach. 

The eigenvalue problem for the Falkner-Skan boundary layer is the following:
\begin{equation}
    N_{k,\eta\eta\eta} + F_\beta N_{k,\eta\eta} + (\lambda_k - 2\beta)F_{\beta,\eta}N_{k,\eta} + (1 - \lambda_k)F_{\beta,\eta\eta}N_k = 0 \label{eq:fs_eigenvalue}
\end{equation} 
with boundary conditions $N_k(0)=0$, $N_{k,\eta}(0)=0$ and $N_{k,\eta} \rightarrow 0$ exponentially as $\eta \rightarrow \infty$. $F_\beta(\eta)$ is the solution to the Falkner-Skan equation: 
\begin{equation}
    F_{\beta,\eta\eta\eta} + F_\beta F_{\beta,\eta\eta} + \beta(1 - F_{\beta,\eta}^2) = 0 \label{eq:fs_equation}
\end{equation} 
with the same boundary conditions as for the flat plate. 
Equation \eqref{eq:fs_equation} corresponds, essentially, to a boundary layer along a flat plate with the outer velocity varying as $U(x) = Kx^m$ (where $m>0$ is associated with a favorable external pressure gradient), or as the boundary layer flow over a wedge of half-angle $\beta\pi/2$, where $m$ and $\beta$ are related as $\beta = 2m/(m+1)$. 
The similarity variable is in this case \cite{White1991}: 
\begin{equation}
    \eta = y \sqrt{\frac{m+1}{2} \frac{U(x)}{\nu x}} \label{eq:fs_similarity_variable}
\end{equation} 
The corresponding system for the Blasius problem is recovered from either  \eqref{eq:fs_eigenvalue} or \eqref{eq:fs_equation} by setting $\beta=0$. 

The eigenvalues and eigenfunctions for this case depend non-trivially on the pressure gradient parameter $\beta$. 
Solutions for physically relevant flows only exist in a limited range $-0.1988 \le \beta \le 2$. 
For $\beta < 0$, the adverse pressure gradient can lead to boundary layer separation. 
As for the perturbation equation, for positive $\beta$ the situation is in all respects as in the Blasius case, with discrete, real and positive eigenvalues and orthogonal eigenfunctions. 
For negative $\beta \ge -0.1988$, the base flow has two solution branches. 
For the upper branch, the eigenvalues are again real and positive, while for the lower branch infinite sets of positive and negative eigenvalues exist. 
In the critical case $\beta = -0.1988$, $F_{\beta,\eta\eta}(0) = 0$ gives rise to separation profiles. 
This case is pathological because $F_{\beta,\eta}$ is an eigenfunction for any eigenvalue. 

The Green's function for this case has been determined by Chen and Libby \cite{ChenLibby1968} as:
\begin{equation}
    G(s',\eta'; s,\eta) = -{\cal H}(s' - s) \sum_{k=1}^\infty D_k(\eta) \frac{1}{2s} \left( \frac{s'}{s} \right)^{-\lambda_k/2} N_k(\eta') \label{eq:fs_greens_function}
\end{equation} 
where $s(x) = U(x)x/\nu$ is the streamwise Levy-Lees variable \cite{ChenLibby1968} and  
\begin{equation}
    D_k(\eta) = \frac{1}{C_k} \frac{w(\eta)}{F_{\beta,\eta}} \left( \frac{N_k}{F_{\beta,\eta}} \right)_\eta \label{eq:fs_adjoint_eigenfunction}
\end{equation} 
are the adjoint eigenfunctions obeying:
\begin{equation}
    -D_{k,\eta\eta\eta} + F_\beta D_{k,\eta\eta} + (2 - \lambda_k + 2\beta)F_{\beta,\eta} D_{k,\eta} + (2 + 2\beta - 2\lambda_k)F_{\beta,\eta\eta} D_k = 0 \label{eq:fs_adjoint_pde}
\end{equation} 
which is adjoint to  \eqref{eq:fs_eigenvalue}. In  \eqref{eq:fs_adjoint_eigenfunction}, 
\begin{equation}
    C_k = \int_0^\infty F_{\beta,\eta} w(\eta) \left( \frac{N_k}{F_{\beta,\eta}} \right)_\eta^2 d\eta \label{eq:fs_norm}
\end{equation} 
are the normalizing constants and 
\begin{equation}
    w(\eta) = F_{\beta,\eta}^3(\eta) e^{\int_0^\eta F_\beta(\eta') d\eta'} \label{eq:fs_weight_function}
\end{equation} 
is the Sturm-Liouville weight. Using \eqref{eq:fs_greens_function}-\eqref{eq:fs_weight_function} we can again work out the corresponding semi-analytic adjoint solution for the integrated friction drag 

\begin{equation}
C_D(L) = \frac{2\nu}{L} \int_0^L \frac{u_y}{U(x)^2} dx
\label{eq:FSdrag}
\end{equation} 

Inserting the Green's function \eqref{eq:fs_greens_function} into (\ref{eq:FSdrag}) and integrating over the Levy-Lees variable $s'$, we recover the $x$-momentum adjoint variable for the Falkner-Skan flow:
\begin{equation}
    \tilde{\psi}(x,\eta) = \frac{2}{L U(x)^2} \sum_{k=1}^\infty \frac{D_k(\eta)}{\lambda_k+\beta-1} \left( 1 - \xi^{\frac{\lambda_k+\beta-1}{2-\beta}} \right) \label{eq:fs_adjoint_momentum}
\end{equation}
where $\xi = x/L$. Notice that \eqref{eq:fs_adjoint_momentum} reduces to the Blasius case \eqref{eq:adjoint_momentum_final} for $\beta=0$. 

The adjoint eigenfunctions also obey the following identity 
\begin{equation}
    \sum_{k=1}^\infty \frac{1 + \beta N_k(\infty)}{\lambda_k - \beta - 1} D_k = 1 \label{eq:fs_eigenfunction_identity}
\end{equation}
which can be obtained by integrating the corresponding adjoint Sturm-Liouville equation. From (\ref{eq:fs_eigenfunction_identity}), the following relationship involving the normalization constants and the eigenvalues can be derived:
\begin{equation}
    \int_0^\infty e^{-\int_0^\eta F_\beta d\eta'} d\eta = \sum_{k=1}^\infty \frac{(1 + \beta N_k(\infty))^2}{C_k (\lambda_k - \beta - 1)^2} \label{eq:fs_norm_identity}
\end{equation}
The integral on the left-hand side \eqref{eq:fs_norm_identity} has the value $1.85536$, providing a constraint for the eigenvalues. 

%
%

\begin{figure}[htbp]
    \centering
    \includegraphics[width=0.7\textwidth]{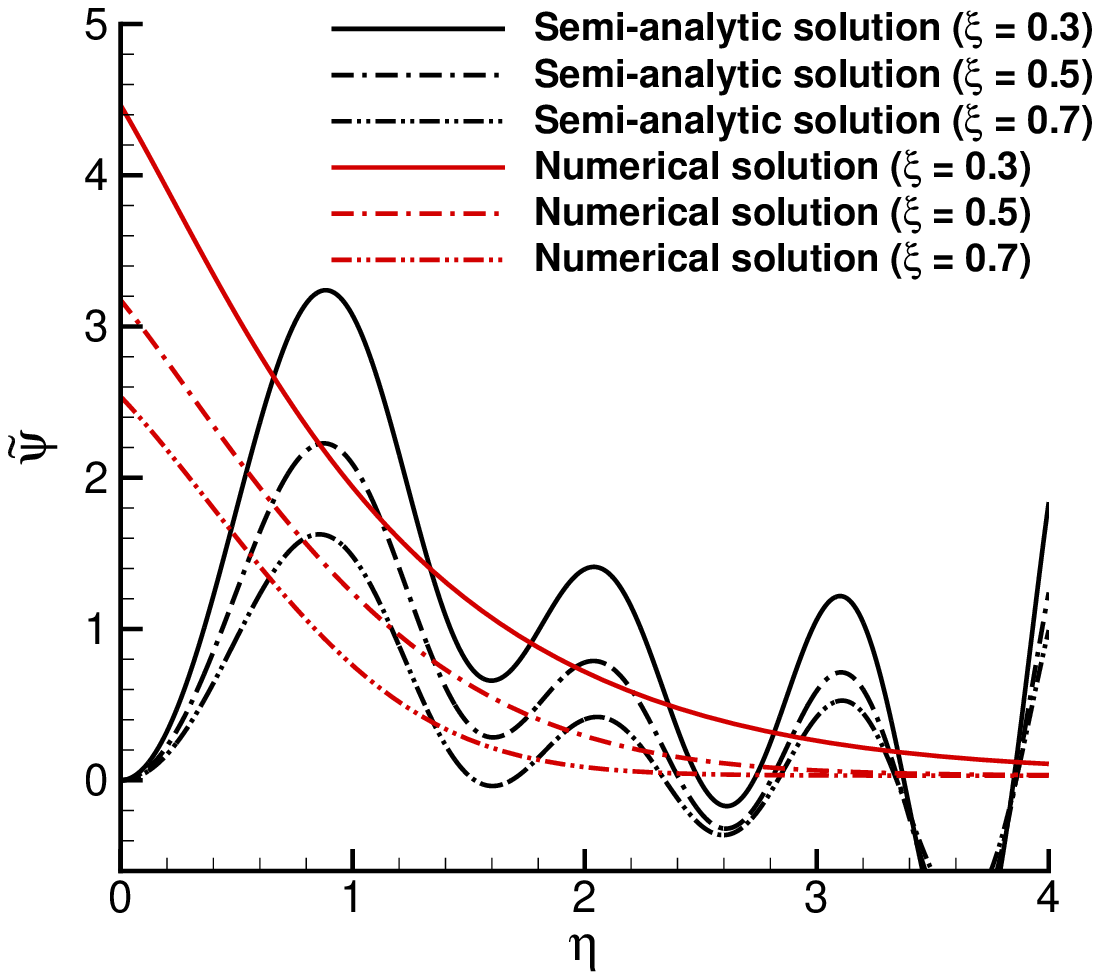}
    \caption{Semi-analytic adjoint solution for Falkner-Skan flow with $\beta=1/2$ computed with the first 20 eigenfunctions described in \cite{ChenLibby1968}.}
    \label{fig:fig6}
\end{figure}

To illustrate the shape of the adjoint solution, we plot in figure \ref{fig:fig6} the result obtained for $\beta=1/2$ using the first 20 eigenfunctions and eigenvalues reported by \cite{ChenLibby1968}, along with results obtained with the finite-volume incompressible  solver described in section \ref{sec:numerical_analysis} .

\begin{figure}[H]
    \centering
    \includegraphics[width=0.7\textwidth]{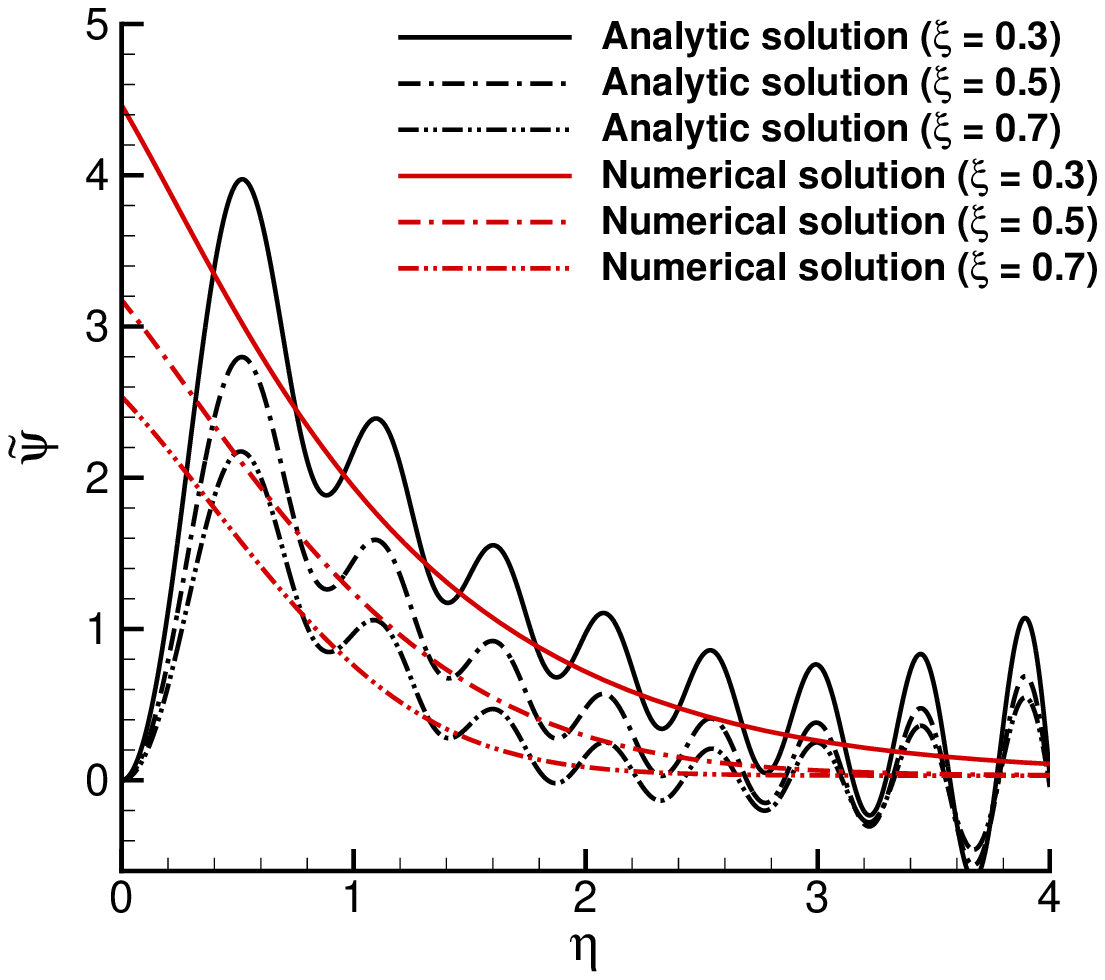}
    \caption{Semi-analytic adjoint solution for Falkner-Skan flow with $\beta=1/2$ computed with 100 eigenfunctions, illustrating the severe Gibbs phenomenon.}
    \label{fig:fig7}
\end{figure}

The agreement with only 20 modes is remarkably poor, and the semi-analytic results are highly oscillatory. This highlights a clear limitation of the eigenfunction expansion approach in the presence of pressure gradients: the convergence is extremely slow. By extrapolating the eigenvalues and norms with the asymptotic fits 
$\lambda_n \sim 3.320 + 2.002(n-1) - 0.2875\sqrt{n-1} - 0.075/\sqrt{n-1}$ and $C_n \sim (1.05 + 1.01/n)n^{-2/3}$,  
we extended the expansion to 100 terms. The results are plotted in figure \ref{fig:fig7}. 

While the 100-mode expansion begins to align on average with the numerical solution, it still shows severe high-frequency oscillations typical of the Gibbs phenomenon. This indicates that, while the solution \eqref{eq:fs_adjoint_momentum} is formally correct, its practical utility for constructing smooth adjoint fields in Falkner-Skan flows is highly limited compared to the zero-pressure-gradient case.

\section{Conclusions}
\label{sec:conclusions}

In this paper, the drag-based adjoint solution for the two-dimensional momentum equation for a laminar boundary layer along a flat plate has been investigated. Using the Green's function approach, the adjoint solution is written as an expansion over the algebraically developing steady disturbance modes described by Libby and Fox. The analysis of this solution has revealed a number of interesting facts, not only about the adjoint equations, but also about Libby-Fox theory.  

First, the solution can be used to analyze the spatial structure of the adjoint field, making it possible to answer questions that have been raised in recent literature. For example, it has recently been claimed that under suitable approximations, the adjoint Blasius solution collapses into a simple backward-facing similarity profile. Our analysis demonstrates that this is not the case. Since the Blasius
base flow possesses a privileged origin at the leading edge, the adjoint equations lack the translation invariance required to sustain a backward-facing similarity variable (a property that is enjoyed by translationally invariant cases such as the Oseen-Blasius flow). 

Second, the solution can be used to analyze the Adjoint Transport Convection (ATC) terms. Frequently cited as a source of numerical instability and sometimes neglected in industrial solvers, our analysis shows that, at least for this simplified setting, the ATC is of the same order of magnitude as the standard adjoint convective terms, so its omission is not theoretically justified. The impact that its omission can have on the sensitivities or on specific applications (such as in optimal aerodynamic shape design) under the present or more complex scenarios is a separate issue that has not been addressed here. 

Third, this formulation offers a new way to address topics covered in the original Libby–Fox study. Applying the adjoint approach to a steady 2D active flow control case involving wall suction and blowing, we recover the solutions obtained by Libby and Fox in 1963. As in the original work, these results assume that the perturbed flow is laminar an remains sufficiently close to the Blasius solution and potential instability issues are ignored. However, the obtained agreement provides support for the present formulation and leads to a number of numerical identities involving the Libby-Fox eigenfunctions and eigenvalues.

Sample solutions have been generated that compare extraordinarily well with numerical adjoint solutions, thus confirming its potential as a benchmarking tool for verification of adjoint solvers.  
The approach has focused on the integrated skin friction, but the Green's function can be used to compute the adjoint solution for any conceivable quantity of interest that is sensitive to changes in the streamwise momentum. 

Finally, we have outlined the analysis for the case of non-zero pressure gradient boundary layers obeying the Falkner-Skan equations. 
While the Green’s function approach successfully yields an exact adjoint solution, the resulting eigenfunction expansion is severely limited by slow convergence issues and the Gibbs phenomenon. Hence, while the eigenfunction expansion is a valid tool for the zero-pressure-gradient Blasius flow, its practical utility does not extend seamlessly to generalized pressure-gradient boundary layers.

There are several possible avenues for future work. Fox and Libby also studied perturbations to the energy equation in Blasius boundary layers, and similar analysis exist concerning the equation of species concentrations. 
The associated eigenvalue problems are somewhat simpler than the one considered here, being inherently second order and linear in the absence of coupling between velocity and energy and/or composition (decoupling that has precluded the application of those results to the computation of the drag-based adjoint energy variable here) and can be used to generate adjoint solutions in those settings. 
Finally, the fact that the adjoint eigenfunctions obey a second order ordinary differential equation (as opposed to Libby-Fox's third order ODE) in which exponential decay of the solutions is replaced by polynomial growth may offer an alternative method for determining the eigenvalues. 
Unfortunately, the equation appears to be poorly conditioned, which has hampered progress along this line. 
Some of these themes will be pursued elsewhere.

\section*{Acknowledgements}
Some of the computations described in this paper were carried out with an incompressible adjoint BL/NS solver built upon the edu2D solver developed by Dr. H. Nishikawa which is freely available at his Researchgate profile https://www.researchgate.net/profile/Hiroaki-Nishikawa-2?ev=brs\verb|_|overview. 

\section*{Funding}
The research described in this paper has been supported by INTA under grant IDATEC (IGB21001). 

\section*{Declaration of Interests}
The authors report no conflict of interest.

\appendix
\begin{appendices}

\section{Explicit adjoint solution for the Oseen-Blasius\\ equation}
\label{app:linear_problem}

Here we apply the Green's function approach to the Oseen linearization of the boundary layer equations (also known as the Oseen-Blasius equation). This equation arises as the first approximation in perturbative approaches to the Blasius solution such as \cite{Kusukawa2014}, but it can also be understood as the limit of the Blasius equation near the edge of the boundary layer. Unlike the Blasius flow, which possesses a privileged origin at the plate's leading edge, the Oseen approximation assumes a constant base velocity profile, rendering the problem translationally invariant. This physical distinction fundamentally alters the mathematical structure of the adjoint solution and allows for a closed-form similarity solution.

The Oseen-Blasius equation is:
\begin{equation}
    (U,0)\cdot\nabla u - \nu \partial_{yy}^2 u = 0 \label{eq:oseen_blasius_pde}
\end{equation}
with $u=0$ at the wall and $u\to U$ as $y\to\infty$. This problem admits the exact self-similar solution $u/U = \text{erf}(\eta/\sqrt{2})$, where $\text{erf}(x)$ is the error function and $\eta$ is the standard Blasius variable. 

The corresponding adjoint equation for the integrated drag objective function is:
\begin{equation}
    (U,0)\cdot\nabla\tilde{\psi} + \nu \partial_{yy}^2 \tilde{\psi} = 0 \label{eq:adjoint_oseen_pde}
\end{equation}
with boundary conditions (\ref{eq:adjoint_boundary_conditions}).   
In the Oseen approximation, the assumption of a uniform background velocity renders the flow translation-invariant. Because the physical system retains no memory of the plate's leading edge, the upstream-propagating adjoint information is governed solely by the distance from the trailing edge, $L-x$. It is precisely this spatial symmetry that allows the adjoint field to collapse into the self-similar solution (\ref{eq:adj_oseen_blasius}) with a profile identical to the primal flow and similarity variable $\hat{\eta} \propto y/\sqrt{L-x}$. 

To verify our Green's function methodology, we can derive the adjoint solution using an eigenfunction expansion. 
The corresponding eigenvalue problem is formally identical to the one described by Chen \cite{Chen1971} for the laminar boundary layer on a continuous moving plate. The eigenvalues are even integers $\lambda_k = 2k$, with $k = 1,2,...$ and the eigenfunctions are $He_{2k-1}(\eta)e^{-\eta^2/2}$, where $He_k(\eta)$ are the Chebyshev-Hermite polynomials. Chen also derived the corresponding Green's function, which can be used to cast the drag-based adjoint solution as the following eigenfunction expansion:
\begin{equation}
    \tilde{\psi}({x},y) = \frac{2}{L U^2} \left( 1-\sum_{k=1}^\infty \frac{1}{2k-1} \frac{(-1)^{k-1} He_{2k-1}({\eta})}{2^{k-3/2}(k-1)!\sqrt{\pi}} \left(\frac{{x}}{L}\right)^{k-1/2}\right) \label{eq:adjoint_oseen_series_final}
\end{equation}
Noting that $\hat{\eta}=\sqrt{x/(L-x)}\eta$, the infinite series on the RHS of \eqref{eq:adjoint_oseen_series_final} is precisely the Taylor expansion of $\text{erf}({\hat{\eta}}/\sqrt{2})$ in powers of $x$ around $x=0$. Thus, the eigenfunction expansion strictly recovers the exact similarity solution (\ref{eq:adj_oseen_blasius}).
This confirms that while the eigenfunction expansion perfectly collapses into a simple self-similar profile for the translation-invariant Oseen flow, it does not do so for the non-linear Blasius flow due to the lack of translation invariance.

\section{Asymptotic behavior of the adjoint eigenfunctions for large $\eta$}
\label{app:asymptotic_large_eta}

To establish the bounding properties of the full adjoint solution, we must determine the asymptotic behavior of the adjoint eigenfunctions $D_k(\eta)$ as $\eta \rightarrow \infty$.

As $\eta \rightarrow \infty$, the base Blasius functions behave asymptotically as $F_0 \rightarrow \eta - c$, where $c \approx 1.21678$. The primal Libby-Fox eigenfunctions behave as $N_k \rightarrow \beta_k$ and $N_{k,\eta} \sim (\eta - c)^{\lambda_k - 1} e^{-(\eta - c)^2 / 2}$ \cite{LibbyFox1963, Kotorynski1968}. 

For sufficiently large $\eta$, substituting these asymptotic base flow values into the adjoint Sturm-Liouville equation \eqref{eq:adjoint_sturm_liouville} yields:
\begin{equation}
D_{k,\eta\eta} - (\eta - c)D_{k,\eta} + (\lambda_k - 1)D_k = 0
\label{eq:asymptotic_adjoint_sturm}
\end{equation}
Equation \eqref{eq:asymptotic_adjoint_sturm} is a variation of Hermite's differential equation. For non-integer eigenvalues $\lambda_k$, the general solution is a linear combination of a Hermite function and a confluent hypergeometric function of the first kind:
\begin{equation}
    D_k(\eta) \sim \kappa_1 H_{\lambda_k - 1}(\eta) + \kappa_2 {_1}F_1 \left( \frac{1-\lambda_k}{2}; \frac{1}{2}; \frac{(\eta-c)^2}{2} \right) \label{eq:D_k_hypergeometric}
\end{equation}
As $\eta \rightarrow \infty$, the Hermite function scales as $(\eta - c)^{\lambda_k - 1}$, while the hypergeometric function contains a severely divergent branch scaling as $(\eta - c)^{-\lambda_k} e^{(\eta - c)^2 / 2}$. To satisfy the physical boundary conditions of the adjoint field, the divergent exponential branch must be discarded ($\kappa_2 = 0$). 

Therefore, the acceptable asymptotic behavior for the adjoint eigenfunctions at the boundary layer edge is strictly polynomial:
\begin{equation}
    D_k(\eta) \sim b_k (\eta - c)^{\lambda_k - 1} \label{eq:D_k_asymptotic_final}
\end{equation}
where $b_k$ are modal constants. This polynomial scaling directly governs the boundedness of the eigenfunction expansion $\varphi(\xi,\eta)$ as discussed in Appendix \ref{app:properties_phi}.

\section{Proof of equation (\ref{eq:D_0_definition})}
\label{app:D_0_is_1}

In order to prove  (\ref{eq:D_0_definition}) we begin by integrating both sides of the adjoint Sturm-Liouville equation \eqref{eq:adjoint_sturm_liouville} over $\eta \in [0,\infty)$:
\begin{equation}
    \left. \frac{F_{0,\eta\eta}}{F_{0,\eta}} D_{k,\eta} \right|_0^\infty = -(\lambda_k - 1) \int_0^\infty F_{0,\eta\eta} D_k d\eta \label{eq:integrated_sturm_liouville}
\end{equation} 
The left-hand side (LHS) is $-1/C_k$ so, rearranging
\begin{equation}
   C_k  \int_0^\infty F_{0,\eta\eta} D_k d\eta = \frac{1}{\lambda_k - 1} \label{eq:integral_D_n}
\end{equation} 
Using the orthogonality properties of the adjoint eigenfunction, it is easy to see that the LHS of (\ref{eq:integral_D_n}) is the coefficient of $D_k$ in the eigenfunction expansion of the constant function 1, so that 
$\Sigma_k D_k/(\lambda_k-1)=1$.

%
%

\subsection{A constraint on the eigenvalues}

Equation \eqref{eq:D_0_definition} can also be used to prove the following identity:
\begin{equation}
    \sum_{n=1}^\infty \frac{1}{C_n (\lambda_n - 1)^2} = 1 \label{eq:eigenvalue_constraint}
\end{equation} 
To show this, consider the following relation that follows from  \eqref{eq:D_0_definition}:
\begin{equation}
    \int_0^\infty F_{0,\eta\eta}(\eta) \left( \sum_{n=1}^\infty \frac{D_n(\eta)}{\lambda_n - 1} \right)^2 d\eta = \int_0^\infty F_{0,\eta\eta}(\eta) d\eta = F_{0,\eta}(\eta) \Big|_0^\infty = 1 \label{eq:identity_integral}
\end{equation} 
On the other hand, expanding the squared term on the LHS of  \eqref{eq:identity_integral} and using the orthogonality relations \eqref{eq:adjoint_orthogonality}, yields:
\begin{equation}
    \int_0^\infty F_{0,\eta\eta} \left( \sum_{n=1}^\infty \frac{D_n}{\lambda_n - 1} \right)^2 d\eta = \sum_{n=1}^\infty \frac{1}{C_n (\lambda_n - 1)^2} \label{eq:lhs_expansion}
\end{equation} 
from where  \eqref{eq:eigenvalue_constraint} readily follows. 
No closed form is known for the eigenvalues and the norms, apart from Brown's asymptotic result \eqref{eq:brown_formula}, but  \eqref{eq:eigenvalue_constraint} is a nice constraint. 

We have checked (\ref{eq:eigenvalue_constraint}) with the values of $\lambda_n$ and $C_n$ determined in section \ref{sec:numerical_analysis}. With 2800 terms, the approximate value is 0.9615. 
%
%

\section{Properties of the eigenfunction expansion $\varphi$}
\label{app:properties_phi}

The semi-analytic adjoint solution is governed by the expansion function $\varphi(\xi,\eta)$ defined in \eqref{eq:phi_definition}. This appendix establishes its rigorous mathematical properties—specifically its asymptotic limits and strict boundedness—which are required to satisfy the global adjoint boundary conditions and to formally bound the expansion.

From \eqref{eq:D_0_definition} and the standard properties of Dirichlet series \cite{Serre1973}, the series defining $\varphi(\xi,\eta)$ converges uniformly for all $\xi \in [0,1]$ and all $\eta$, yielding an analytic function. Furthermore, we can use the Sturm-Liouville orthogonality \eqref{eq:lhs_expansion} to strictly bound the norm of $\varphi$:
\begin{equation}
\begin{aligned}
    ||\varphi||^2 &= \int_0^\infty F_{0,\eta\eta} \varphi^2 d\eta = \int_0^\infty F_{0,\eta\eta} \sum_{k,j=1}^\infty \frac{D_k(\eta)D_j(\eta)}{(\lambda_k - 1)(\lambda_j - 1)} \xi^{\frac{\lambda_k - 1}{2}} \xi^{\frac{\lambda_j - 1}{2}} d\eta \\
    &= \sum_{k=1}^\infty \frac{1}{C_k(\lambda_k - 1)^2} \xi^{\lambda_k - 1} \le \sum_{k=1}^\infty \frac{1}{C_k(\lambda_k - 1)^2} = 1
\end{aligned} \label{eq:phi_norm_bound}
\end{equation}

\subsection{Asymptotic behavior for large $\eta$}

As demonstrated in Appendix \ref{app:asymptotic_large_eta}, the adjoint eigenfunctions scale asymptotically as $D_k \sim b_k (\eta - c)^{\lambda_k - 1}$ for large $\eta$. Substituting this into the expansion yields the asymptotic behavior of $\varphi$:
\begin{equation}
    \varphi(\xi,\eta) \sim \sum_{k=1}^\infty \frac{b_k}{\lambda_k - 1} \left( \xi^{1/2}(\eta - c) \right)^{\lambda_k - 1} \equiv f\big(\xi^{1/2}(\eta - c)\big) \label{eq:phi_asymptotic_sum}
\end{equation}
where $f(z)$ represents the infinite series in terms of the composite variable $z = \xi^{1/2}(\eta - c)$. 

Evaluating this at $\xi = 1$ yields $\varphi(1,\eta) \sim f(\eta - c)$. Since $\varphi(1,\eta)=1$ for all $\eta$, the function $f(z)$ must approach $1$ as $z \rightarrow \infty$. Consequently, for any fixed $\xi > 0$:
\begin{equation}
    \lim_{\eta \rightarrow \infty} \varphi(\xi,\eta) = \lim_{\eta \rightarrow \infty} f\big(\xi^{1/2}(\eta - c)\big) = 1 \label{eq:limit_phi}
\end{equation}
Inserting this limit back into the definition of the adjoint variable \eqref{eq:analytic_adjoint_drag} proves that $\lim_{\eta \rightarrow \infty} \tilde{\Psi}(x,\eta) = 0$, as expected.

\subsection{Boundedness via the Maximum Principle}

Numerical evaluations in Section \ref{sec:numerical_analysis} suggested that $0 \le \varphi \le 1$. We can prove this analytically by showing that the eigenfunction expansion obeys a difussion equation and invoking the Maximum Principle \citep{Evans2010}. 

Consider the individual terms of the expansion, $T_n(\xi,\eta) = D_n(\eta) \xi^{(\lambda_n - 1)/2} / (\lambda_n - 1)$. Introducing a pseudo-time variable $t = -\frac{1}{2}\ln\xi$ to map the domain $\xi \in (0,1]$ to $t \in [0,\infty)$, we obtain $\partial_t T_n = -(\lambda_n - 1) T_n$, which combined with the Sturm-Liouville equation \eqref{eq:adjoint_sturm_liouville} yields:
\begin{equation}
    \partial_t T_n = \frac{1}{F_{0,\eta\eta}} \left( \frac{F_{0,\eta\eta}}{F_{0,\eta}} T_{n,\eta} \right)_\eta \label{eq:T_n_t_derivative}
\end{equation}
Defining $z = F_{0,\eta}(\eta)$ maps the semi-infinite physical domain $\eta \in [0,\infty)$ to a bounded domain $z \in [0,1]$. Under this transformation, \eqref{eq:T_n_t_derivative} simplifies to a standard diffusion equation $\partial_t T_n = \partial_z \left( \kappa(z) \partial_z T_n \right)$ with a variable diffusivity $\kappa(z) = F_{0,\eta\eta}^2 / z$. 

Summing over all modes $n$ shows that $\varphi$ obeys the same diffusion equation:
\begin{equation}
    \partial_t \varphi = \partial_z \left( \kappa(z) \partial_z \varphi \right) \label{eq:phi_heat_equation}
\end{equation}
defined on the bounded domain $z \in [0,1]$ and $t \in [0,\infty)$ and subject to the conditions $\varphi(t=0,z) = 1$, $\varphi(t\to\infty,z) = 0$, $\varphi(t,0) = 0$, and $\varphi(t,1) = 1$. 

The effective diffusivity $\kappa(z)$ is strictly positive in the interior of the domain, so the strong Maximum Principle for parabolic PDEs applies \citep{Evans2010}, according to which the maximum and minimum values of $\varphi$ cannot occur in the interior of the domain; they must occur either on the spatial boundaries or at the initial state $t=0$. Hence,
\begin{equation}
    0 \le \varphi(\xi,\eta) \le 1
\end{equation}

\end{appendices}


\bibliographystyle{unsrt} 
\bibliography{references} 

\end{document}